\newif\ifdraft\draftfalse
\newif\iffull\fullfalse
\newif\ifcolor\colortrue

\makeatletter \@input{texdirectives} \makeatother

\documentclass[nocopyrightspace]{sigplanconf}

\usepackage{pervasives}

\toappear{}

\title{A Fast Compiler for NetKAT}

\authorinfo{Steffen Smolka}{Cornell University (USA)}{smolka@cs.cornell.edu}
\authorinfo{Spiridon Eliopoulos\thanks{Work performed at Cornell University.}}{Inhabited Type LLC (USA)}{spiros@inhabitedtype.com}
\authorinfo{Nate Foster}{Cornell University (USA)}{jnfoster@cs.cornell.edu}
\authorinfo{Arjun Guha}{UMass Amherst (USA)}{arjun@cs.umass.edu}

\begin{document}

\maketitle

\begin{abstract}
High-level programming languages play a key role in a growing number
of networking platforms, streamlining application development and
enabling precise formal reasoning about network behavior.
Unfortunately, current compilers only handle ``local'' programs that
specify behavior in terms of hop-by-hop forwarding behavior, or modest
extensions such as simple paths. To encode richer ``global''
behaviors, programmers must add extra state---something that is tricky
to get right and makes programs harder to write and maintain. Making
matters worse, existing compilers can take tens of minutes to generate
the forwarding state for the network, even on relatively small
inputs. This forces programmers to waste time working around
performance issues or even revert to using hardware-level \textsc{api}s.

This paper presents a new compiler for the \netkat language that
handles rich features including regular paths and virtual networks,
and yet is several orders of magnitude faster than previous compilers. The
compiler uses symbolic automata to calculate the extra state needed to
implement ``global'' programs, and an intermediate representation
based on binary decision diagrams to dramatically improve
performance. We describe the design and implementation of three
essential compiler stages: from virtual programs (which specify
behavior in terms of virtual topologies) to global programs (which
specify network-wide behavior in terms of physical topologies), from
global programs to local programs (which specify behavior in terms of
single-switch behavior), and from local programs to hardware-level
forwarding tables. We present results from experiments on real-world
benchmarks that quantify performance in terms of compilation time and
forwarding table size.
\end{abstract}

\category{D.3.4}
{Programming Languages}
{Processors}
[Compilers]

\keywords
Software-defined networking,
domain-specific languages,
Net\textsc{kat},
Frenetic,
Kleene Algebra with tests,
virtualization,
binary decision diagrams.

\section{Introduction}

High-level languages are playing a key role in a growing number of
networking platforms being developed in academia and industry. There
are many examples: VMware uses \texttt{nlog}, a declarative language
based on Datalog, to implement network
virtualization~\cite{vmware:nvp}; \textsc{sdx} uses Pyretic to combine
programs provided by different participants at Internet exchange
points~\cite{gupta:sdx,monsanto:pyretic}; \textsc{pane} uses NetCore
to allow end-hosts to participate in network management
decisions~\cite{ferguson:pane,monsanto:netcore}; Flowlog offers
tierless abstractions based on Datalog~\cite{nelson:flowlog}; Maple
allows packet-processing functions to be specified directly in Haskell
or Java~\cite{voellmy:maple}; OpenDaylight's group-based policies
describe the state of the network in terms of application-level
connectivity requirements~\cite{odl:group}; and \textsc{onos} provides
an ``intent framework'' that encodes constraints on end-to-end
paths~\cite{onos:intent}.

The details of these languages differ, but they all offer abstractions
that enable thinking about the behavior of a network in terms of
high-level constructs such as packet-processing functions rather than
low-level switch configurations. To bridge the gap between these
abstractions and the underlying hardware, the compilers for these
languages map source programs into forwarding rules that can be
installed in the hardware tables maintained by software-defined
networking (\textsc{sdn}) switches.

Unfortunately, most compilers for \textsc{sdn} languages only handle
``local'' programs in which the intended behavior of the network is
specified in terms of hop-by-hop processing on individual switches. A
few support richer features such as end-to-end paths and network
virtualization~\cite{voellmy:maple,onos:intent,vmware:nvp}, but to the
best of our knowledge, no prior work has presented a complete
description of the algorithms one would use to generate the forwarding
state needed to implement these features. For example,
although \netkat includes primitives that can be used to succinctly
specify global behaviors including regular paths, the existing
compiler only handles a local
fragment~\cite{anderson:netkat}. This means that programmers can only
use a restricted subset that is strictly less expressive than the full
language and must manually manage the state needed to implement
network-wide paths, virtual networks, and other similar features.

Another limitation of current compilers is that they are based on
algorithms that perform poorly at scale. For example, the
NetCore, \netkat, \textsc{pane}, and Pyretic compilers use a simple
translation to forwarding tables, where primitive constructs are
mapped directly to small tables and other constructs are mapped to
algebraic operators on forwarding tables. This approach quickly
becomes impractical as the size of the generated tables can grow
exponentially with the size of the program!  This is a problem for
platforms that rely on high-level languages to express control
application logic, as a slow compiler can hinder the ability of the
platform to effectively monitor and react to changing network state.

Indeed, to work around the performance issues in the current Pyretic
compiler, the developers of \textsc{sdx}~\cite{gupta:sdx} extended the
language in several ways, including adding a new low-cost composition
operator that implements the disjoint union of packet-processing
functions. The idea was that the implementation of the disjoint union
operator could use a linear algorithm that simply concatenates the
forwarding tables for each function rather than using the usual
quadratic algorithm that does an all-pairs intersection between the
entries in each table. However, even with this and other
optimizations, the Pyretic compiler still took tens of minutes to
generate the forwarding state for inputs of modest size.

\paragraph*{Our approach.} 
This paper presents a new compiler pipeline for \netkat that handles
local programs executing on a single switch, global programs that
utilize the full expressive power of the language, and even programs
written against virtual topologies. The algorithms that make up this
pipeline are orders of magnitude faster than previous
approaches---e.g., our system takes two seconds to compile the
largest \textsc{sdx} benchmarks, versus several minutes in Pyretic,
and other benchmarks demonstrate that our compiler is able to handle
large inputs far beyond the scope of its competitors.

These results stem from a few key insights. First, to compile local
programs, we exploit a novel intermediate representation based on
binary decision diagrams (\bdds). This representation avoids the
combinatorial explosion inherent in approaches based on forwarding
tables and allows our compiler to leverage well-known techniques for
representing and transforming \bdds. Second, to compile global
programs, we use a generalization of symbolic
automata~\cite{pous:symbolickat} to handle the difficult task of
generating the state needed to correctly implement features such as
regular forwarding paths. Third, to compile virtual programs, we
exploit the additional expressiveness provided by the global compiler
to translate programs on a virtual topology into programs on the
underlying physical topology.

We have built a full working implementation of our compiler in OCaml,
and designed optimizations that reduce compilation time and the size
of the generated forwarding tables. These optimizations are based on general
insights related to \bdds (sharing common structures, rewriting naive
recursive algorithms using dynamic programming, using heuristic field
orderings, etc.) as well as domain-specific insights specific
to \textsc{sdn} (algebraic optimization of \netkat programs,
per-switch specialization, etc.). To evaluate the performance of our
compiler, we present results from experiments run on a variety of
benchmarks. These experiments demonstrate that our compiler provides
improved performance, scales to networks with tens of thousands of
switches, and easily handles complex features such as virtualization.

Overall, this paper makes the following contributions:
\begin{itemize}
\item We present the first complete compiler pipeline for \netkat that
translates local, global, and virtual programs into forwarding tables
for \textsc{sdn} switches.

\item We develop a generalization of \bdds and show how to implement a local
 \textsc{sdn} compiler using this data structure as an intermediate
 representation.

\item We describe compilation algorithms for virtual and global 
programs based on graph algorithms and symbolic automata.

\item We discuss an implementation in OCaml and develop optimizations 
that reduce running time and the size of the generated forwarding tables.

\item We conduct experiments that show dramatic improvements over
 other compilers on a collection of benchmarks and case studies.
\end{itemize}
The next section briefly reviews the \netkat language and discusses
some challenges related to compiling \textsc{sdn} programs, to set the
stage for the results described in the following sections.

\begin{figure}
\noindent
\center
\yellowbox{
\(
\begin{array}{@{\quad}c@{\quad}}
\multicolumn{1}{@{\quad}l}{\textbf{Syntax}}\\[.5em]
\begin{array}{r@{~~}r@{~}c@{~}l@{\qquad\qquad}l}
\textrm{Naturals} & n & \Coloneqq  & 0 \mid 1 \mid 2 \mid \dots\\[.125em]
\textrm{Fields} & \field & \Coloneqq  & \multicolumn{2}{l}{\field_1 \mid \dots \mid \field_k} \\[.125em]
\textrm{Packets} & \pk & \Coloneqq & \multicolumn{2}{l}{\set{\field_1=n_1, \cdots , \field_k = n_k}} \\[.125em]
\textrm{Histories} & \h & \Coloneqq & \multicolumn{2}{l}{\hone{\pk} \mid \hcons{\pk}{\h}} \\[.125em]
\textrm{Predicates} & \preds &
   \Coloneqq & \ptrue                  & \textit{Identity} \\
    & & \mid & \pfalse                 & \textit{Drop} \\
    & & \mid & \match{\field}{n}       & \textit{Test} \\
    & & \mid & \punion{\preda}{\predb} & \textit{Disjunction} \\
    & & \mid & \pseq{\preda}{\predb}   & \textit{Conjunction} \\
    & & \mid & \pnot{\preda}           & \textit{Negation} \\[.125em]
\textrm{Programs} & \pols &
  \Coloneqq & a                        & \textit{Filter} \\
    & & \mid & \modify{\field}{n}      & \textit{Modification} \\
    & & \mid & \punion{\polp}{\polq}   & \textit{Union} \\
    & & \mid & \pseq{\polp}{\polq}     & \textit{Sequencing} \\
    & & \mid & \pstar{\polp}           & \textit{Iteration} \\
    & & \mid & \pdup                   & \textit{Duplication}
\end{array}\\
\ \\\hline
\ \\
\multicolumn{1}{@{\quad}l}{\textbf{Semantics}}\\[.5em]
\begin{array}{r@{~}c@{~}ll}
\multicolumn{4}{l}{\den{\polp} \in \Hist \rightarrow \powerset{\Hist}}\\[.25em]
\den{\ptrue}~\h & \defeq &
  \set{\h} & \\[.25em]
\den{\pfalse}~\h & \defeq &
  \set{} & \\[.25em]
\den{\match{\field}{n}}~(\hcons{\pk}{\h}) & \defeq &
  \left\{ \begin{array}{@{}ll}
    \set{\hcons{\pk}{\h}} & \textrm{if}~\pkproj{\pk}{\field} = n \\
    \set{} & \textrm{otherwise}\\
  \end{array} \right.\\[.5em]
\den{\pnot\preda}~\h & \defeq &
  \set{\h} \setminus (\den{\preda}~\h) \\[.25em]
\den{\modify{\field}{n}}~(\hcons{\pk}{\h}) & \defeq &
  \set{\hcons{\pkupd{\pk}{\field}{n}}{\h}} \\[.25em]
\den{\punion{\polp}{\polq}}~\h & \defeq &
  \den{\polp}~\h \cup \den{\polq}~\h \\[.25em]
\den{\pseq{\polp}{\polq}}~\h & \defeq &
  (\den{\polp} \Kleisli \den{\polq})~\h \\[.25em]
\den{\pstar\polp}~\h & \defeq &
  \bigcup_i F^i~\h \\
\multicolumn{3}{r}{\text{where}~F^0~\h \defeq \{\h\}~\text{and}~F^{i+1}~\h~\defeq~(\den{\polp} \Kleisli F^i)~\h}\\[.25em]
\den{\pdup}~(\hcons{\pk}{\h}) & \defeq &
  \set{\hcons{\pk}{(\hcons{\pk}{\h})}}\\[.25em]
\end{array}
\end{array}
\)
}
\caption{\netkat syntax and semantics.}
\label{fig:netkat}
\end{figure}

\section{Overview\label{tablecompiler}}
\label{sec:overview}

\netkat is a domain-specific language for specifying and
reasoning about networks~\cite{anderson:netkat,netkat-automata}. It
offers primitives for matching and modifying packet headers, as well
combinators such as union and sequential composition that merge
smaller programs into larger ones. \netkat is based on a solid
mathematical foundation, Kleene Algebra with Tests (\textsc{kat})~\cite{kat},
and comes equipped with an equational reasoning system that can be
used to automatically verify many properties of programs~\cite{netkat-automata}.

\netkat enables programmers to think in terms of functions on packets 
histories, where a packet (\pk{}) is a record of fields and a history
(\h) is a non-empty list of packets. This is a dramatic departure from
hardware-level \textsc{api}s such as OpenFlow, which require thinking
about low-level details such as forwarding table rules, matches,
priorities, actions, timeouts, etc. \netkat fields $f$ include
standard packet headers such as Ethernet source and destination
addresses, \textsc{vlan} tags, \etc, as well as special fields to
indicate the port (\ptf) and switch (\swf) where the packet is located
in the network. For brevity, we use \srcf{} and \dstf{} fields in
examples, though our compiler implements all of the standard fields
supported by OpenFlow~\cite{openflow}.

\paragraph*{\netkat syntax and semantics.}
Formally, \netkat is defined by the syntax and semantics given in
Figure~\ref{fig:netkat}. Predicates $a$ describe logical predicates on
packets and include primitive tests $\match{f}{n}$, which check
whether field $f$ is equal to $n$, as well as the standard collection
of boolean operators. This paper focuses on tests that match fields
exactly, although our implementation supports generalized tests, such
as \textsc{ip} prefix matches. Programs $p$ can be understood as
packet-processing functions that consume a packet history and produce
a set of packet histories. Filters $a$ drop packets that do not
satisfy $a$; modifications $\modify{f}{n}$ update the $f$ field to
$n$; unions $\punion{p}{q}$ copy the input packet and process one copy
using $p$, the other copy using $q$, and take the union of the
results; sequences $\pseq{p}{q}$ process the input packet using $p$
and then feed each output of $p$ into $q$ (the $\bullet$ operator is
Kleisli composition); iterations $\pstar{p}$ behave like the union of
$p$ composed with itself zero or more times; and $\pdup$s extend the
trajectory recorded in the packet history by one hop.

\paragraph*{Topology encoding.}
Readers who are familiar with Frenetic~\cite{foster11},
Pyretic~\cite{monsanto:pyretic}, or NetCore~\cite{monsanto:netcore},
will be familiar with the basic details of this functional
packet-processing model. However, unlike these languages, \netkat can
also model the behavior of the entire network, including its topology.
For example, a (unidirectional) link from port $\pt_1$ on switch $\sw_1$ to port
$\pt_2$ on switch $\sw_2$, can be encoded in \netkat as follows:
\[
\Pseq{\pdup,\match{\swf}{\sw_1},\match{\ptf}{\pt_1},
      \modify{\swf}{\sw_2},\modify{\ptf}{\pt_2},\pdup}
\]
Applying this pattern, the entire topology can be encoded as a union of links. Throughout
this paper, we will use the shorthand
$\plink{\sw_1}{\pt_1}{\sw_2}{\pt_2}$ to indicate links, and assume
that $\pdup$ and modifications to the switch field occur only in
links.

\paragraph*{Local programs.}
Since \netkat can encode both the network topology and the behavior of
switches, a \netkat program describes the end-to-end behavior of a
network. One simple way to write \netkat programs is to define
predicates that describe where packets enter ($\mathit{in}$) and exit
($\mathit{out})$ the network, and interleave steps of processing on
switches ($p$) and topology ($t$):
\[
\small
\Pseq{\textit{in}, \pstar{(\pseq{p}{t})}, p, \textit{out}}
\]
To execute the program, only $p$ needs to be specified---the physical
topology implements $\textit{in}$, $\textit{t}$, and
$\textit{out}$. Because no switch modifications or $\pdup$s occur in
$p$, it can be directly compiled to a collection of forwarding tables,
one for each switch.  Provided the physical topology is faithful to
the encoding specified by $\textit{in}$, $t$, and $\textit{out}$, a
network of switches populated with these forwarding tables will behave
like the above program. We call such a switch program $p$
a \emph{local} program because it describes the behavior of the
network in terms of hop-by-hop forwarding steps on individual
switches.

\begin{figure*}
\begin{subfigure}{0.33\textwidth}
\[
\small
\begin{flowtable}
\star  & \modify{\ptf}{2}
\end{flowtable}
\quad
\mathit{pol}_A\defeq\modify{\ptf}{2}
\]
\caption{An atomic modification}
\end{subfigure}
\begin{subfigure}{0.33\textwidth}
\[
\small
\begin{flowtable}
\match{\dstip}{\ipa}  & \ptrue \\
\star                       & \pfalse
\end{flowtable}
\quad
\mathit{pol}_B\defeq\match{\dstip}{\ipa}
\]
\caption{An atomic predicate}
\end{subfigure}
\begin{subfigure}{0.33\textwidth}
\[
\small
\begin{flowtable}
\match{\dstip}{\ipa}  & \modify{\ptf}{2} \\
\star                       & \pfalse
\end{flowtable}
\quad
\pseq{\mathit{pol}_B}{\mathit{pol}_A}
\]
\caption{Forwarding to a single host}
\end{subfigure}

\begin{subfigure}{0.5\textwidth}
\[
\small
\begin{flowtable}
\match{\dstip}{\ipa}  & \modify{\ptf}{1} \\
\match{\dstip}{\ipb}  & \modify{\ptf}{2} \\
\star                       & \pfalse
\end{flowtable}
\quad
\mathit{pol}_D \defeq
\left.\begin{aligned}
\pseq{\match{\dstip}{\ipa}}{\modify{\ptf}{1}} & \punion{}{} \\
\pseq{\match{\dstip}{\ipb}}{\modify{\ptf}{2}}
\end{aligned}
\right.\]
\caption{Forwarding traffic to two hosts}
\end{subfigure}
\begin{subfigure}{0.5\textwidth}
\[
\small
\begin{flowtable}
\match{\dstip}{\ipa}  & \modify{\ptf}{3} \\
\match{\proto}{\kw{ssh}}  & \modify{\ptf}{3} \\
\star                       & \pfalse
\end{flowtable}
\quad
\mathit{pol}_E \defeq
\pseq{\left(\begin{aligned}
\match{\proto}{\kw{ssh}} & \punion{}{} \\
\match{\dstip}{\ipa}
\end{aligned}
\right)}{\modify{\ptf}{3}}
\]
\caption{Monitoring \textsc{ssh} traffic and traffic to host \ipa}
\end{subfigure}
\smallskip
\caption{Compiling using forwarding tables.}
\label{fig:old-compiler}
\end{figure*}

\paragraph*{Global programs.}
Because \netkat is based on Kleene algebra, it includes regular
expressions, which are a natural and expressive formalism for
describing paths through a network. Ideally, programmers would be able
to use regular expressions to construct forwarding paths directly,
without having to worry about how those paths were implemented. For
example, a programmer might write the following to forward packets
from port $1$ on switch $\sw_1$ to port $1$ on switch $\sw_2$, and
from port $2$ on $\sw_1$ to port $2$ on $\sw_2$, assuming a link
connecting the two switches on port $3$:
\[
\begin{array}{l@{~}l}
  & \Pseq{\match{\ptf}{1},\modify{\ptf}{3},
          \plink{\sw_1}{\pt_3}{\sw_2}{\pt_3},\modify{\ptf}{1}}\\
+ & \Pseq{\match{\ptf}{2},\modify{\ptf}{3},
          \plink{\sw_1}{\pt_3}{\sw_2}{\pt_3},\modify{\ptf}{2}}
\end{array}
\]
Note that this is \emph{not} a local program, since is not written in
the general form given above and instead combines switch processing
and topology processing using a particular combination of union and
sequential composition to describe a pair of overlapping forwarding
paths. To express the same behavior as a local \netkat program or in a
language such as Pyretic, we would have to somehow write a single
program that specifies the processing that should be done at each
intermediate step. The challenge is that when $\sw_2$ receives a
packet from $\sw_1$, it needs to determine if that packet originated
at port $1$ or $2$ of $\sw_1$, but this can't be done without extra
information. For example, the compiler could add a tag to packets at
$\sw_1$ to track the original ingress and use this information to
determine the processing at $\sw_2$. In general, the expressiveness
of \emph{global} programs creates challenges for the compiler, which
must generate explicit code to create and manipulate tags. These
challenges have not been met in previous work on \netkat or
other \textsc{sdn} languages.

\begin{figure}[t]
\includegraphics[width=\columnwidth]{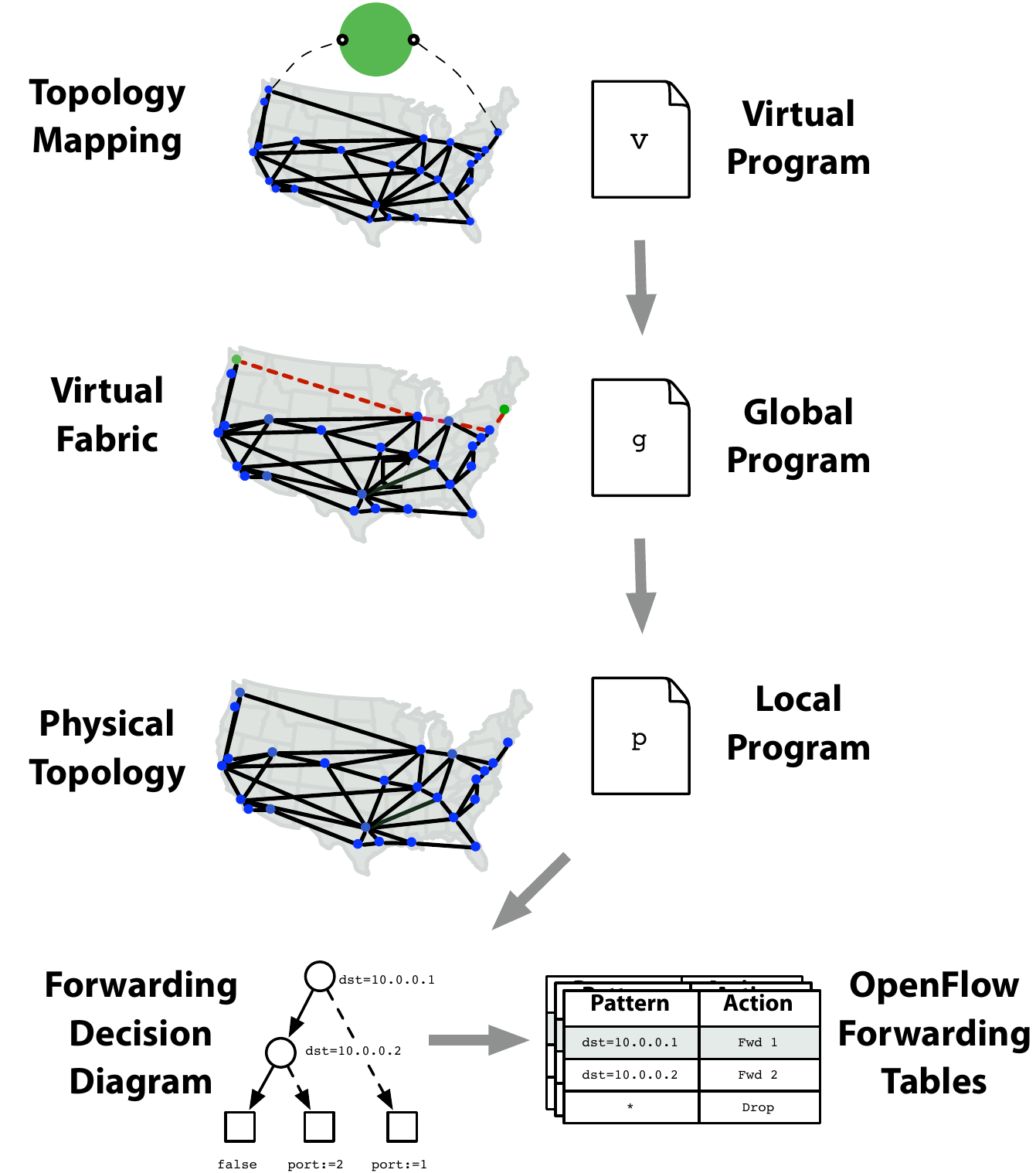}
\caption{\netkat compiler pipeline.}
\label{fig:architecture}
\end{figure}

\paragraph*{Virtual programs.}
Going a step further, \netkat can also be used to specify behavior in
terms of virtual topologies. To see why this is a
useful abstraction, suppose that we wish to implement point-to-point connectivity
between a given pair of hosts in a network with dozens of
switches. One could write a global program that explicitly forwards
along the path between these hosts. But this would be tedious for the
programmer, since they would have to enumerate all of the intermediate
switches along the path.  A better approach is to express the program
in terms of a virtual ``big switch'' topology whose ports are directly
connected to the hosts, and where the relationship between ports in
the virtual and physical networks is specified by an explicit
mapping---\eg, the top of Figure~\ref{fig:architecture} depicts a big
switch virtual topology. The desired functionality could then be
specified using a simple local program that forwards in both
directions between ports on the single virtual switch:
\[
\polp \defeq \punion{(\pseq{\match{\ptf}{1}}{\modify{\ptf}{2}})}
                    {(\pseq{\match{\ptf}{2}}{\modify{\ptf}{1}})}
\]
This one-switch virtual program is evidently much easier to write than
a program that has to reference dozens of switches. In addition, the
program is robust to changes in the underlying network. If the
operator adds new switches to the network or removes switches for
maintenance, the program remains valid and does not need to be
rewritten. In fact, this program could be ported to a completely
different physical network too, provided it is able to implement the
same virtual topology.

Another feature of virtualization is that the physical-virtual mapping
can limit access to certain switches, ports, and even packets that
match certain predicates, providing a simple form of language-based
isolation~\cite{splendid-isolation}.  In this example, suppose the
physical network has hundreds of connected hosts. Yet, since the
virtual-physical mapping only exposes two ports, the abstraction
guarantees that the virtual
program is isolated from the hosts connected to the other ports.
Moreover, we can run several isolated virtual networks on the same
physical network, \eg, to provide different services to different
customers in multi-tenant datacenters~\cite{vmware:nvp}.

Of course, while virtual programs are a powerful abstraction, they
create additional challenges for the compiler since it must generate
physical paths that implement forwarding between virtual ports and
also instrument programs with extra bookkeeping information to keep
track of the locations of virtual packets traversing the physical
network. Although virtualization has been extensively studied
in the networking
community~\cite{monsanto:pyretic,vmware:nvp,casado:presto,shabibi:ovx},
no previous work fully describes how to compile virtual programs.

\paragraph*{Compilation pipeline.}
This paper presents new algorithms for compiling \netkat that address
the key challenges related to expressiveness and performance just
discussed. Figure~\ref{fig:architecture} depicts the overall
architecture of our compiler, which is structured as a pipeline with
several smaller stages: (i) a \emph{virtual compiler} that takes as
input a virtual program $v$, a virtual topology, and a mapping that
specifies the relationship between the virtual and physical topology,
and emits a global program that uses a fabric to transit between
virtual ports using physical paths; (ii) a \emph{global compiler} that
takes an arbitrary \netkat program $g$ as input and emits a local
program that has been instrumented with extra state to keep track of
the execution of the global program; and a (iii) \emph{local compiler}
that takes a local program $p$ as input and generates OpenFlow
forwarding tables, using a generalization of binary decision diagrams
as an intermediate representation. Overall, our compiler automatically
generates the extra state needed to implement virtual and global
programs, with performance that is dramatically faster than
current \textsc{sdn} compilers.

These three stages are designed to work well together---e.g.,
the fabric constructed by the virtual compiler is expressed in terms
of regular paths, which are translated to local programs by the global
compiler, and the local and global compilers both use \fdds as an
intermediate representation. However, the individual compiler stages
can also be used independently. For example, the global compiler
provides a general mechanism for compiling forwarding paths specified
using regular expressions to \textsc{sdn} switches. We have also been
working with the developers of Pyretic to improve performance by
retargeting its backend to use our local compiler.

The next few sections present these stages in detail, starting with
local compilation and building up to global and virtual compilation.

\section{Local Compilation}

The foundation of our compiler pipeline is a translation that maps
local \netkat programs to OpenFlow forwarding tables. Recall that a
local program describes the hop-by-hop behavior of individual
switches---i.e. it does not contain $\pdup$ or switch
modifications.

\paragraph*{Compilation via forwarding tables.}
A simple approach to compiling local programs is to define a
translation that maps primitive constructs to forwarding tables and
operators such as union and sequential composition to functions that
implement the analogous operations on tables. For example, the
current \netkat compiler translates the modification \modify{\ptf}{2}
to a forwarding table with a single rule that sets the port of all
packets to $2$ (Figure~\ref{fig:old-compiler}~(a)), while it
translates the predicate \match{\dstip}{\ipa} to a flow table with two
rules: the first matches packets where \match{\dstip}{\ipa} and leaves
them unchanged and the second matches all other packets and drops them
(Figure~\ref{fig:old-compiler}~(b)).

To compile the sequential composition of these programs, the compiler
combines each row in the first table with the entire second table,
retaining rules that could apply to packets produced by the row
(Figure~\ref{fig:old-compiler}~(c)). In the example, the second table
has a single rule that sends all packets to port $2$. The first rule
of the first table matches packets with destination \ipa, thus the
second table is transformed to only send packets with
destination \ipa{} to port $2$. However, the second rule of the first
table drops all packets, therefore no packets ever reach the second
table from this rule.

To compile a union, the compiler computes the pairwise intersection of
all patterns to account for packets that may match both tables. For
example, in Figure~\ref{fig:old-compiler}~(d), the two sub-programs
forward traffic to hosts \ipa{} and \ipb{} based on the \dstip
header. These two sub-programs do not overlap with each other, which
is why the table in the figure appears simple. However, in general,
the two programs may overlap. Consider compiling the union of the
forwarding program, in Figure~\ref{fig:old-compiler}~(d) and the
monitoring program in Figure~\ref{fig:old-compiler}~(e). The monitoring
program sends \textsc{ssh} packets and packets with
\match{\dstip}{\ipa} to port $3$. The intersection will need to consider
all interactions between pairs of rules---an $\mathcal{O}(n^2)$
operation. Since a \netkat program may be built out of several nested
programs and compilation is quadratic at each step, we can easily get
a tower of squares or exponential behavior.

Approaches based on flow tables are attractive for their simplicity,
but they suffer several serious limitations. One issue is that tables
are not an efficient way to represent packet-processing functions
since each rule in a table can only encode positive tests on packet
headers. In general, the compiler must emit sequences of prioritized
rules to encode operators such as negation or union. Moreover, the
algorithms that implement these operators are worst-case quadratic,
which can cause the compiler to become a bottleneck on large
inputs. Another issue is that there are generally many equivalent ways
to encode the same packet-processing function as a forwarding
table. This means that a straightforward computation of fixed-points,
as is needed to implement Kleene star, is not guaranteed to terminate.

\tikzFdd
\begin{figure}
\begin{subfigure}[t]{0.2\textwidth}
\begin{tikzpicture}

  \node[node] (proto) at (6,-1)
    [label=right:\small\match{\proto{}}{\hdrval{http}}] {};
  \node[node] (ip1) at (5,0)
    [label=right:\small\match{\dstip{}}{\hdrval{10.0.0.1}}] {};
  \node[node] (ip2) at (6,1)
    [label=right:\small\match{\dstip{}}{\hdrval{10.0.0.2}}] {};

  \node[leaf] (pt1) at (5,2)
    [label=below:\small\modify{\ptf{}}{\hdrval{1}}] {};
  \node[leaf] (pt2) at (6,2)
    [label=below:\small\modify{\ptf{}}{\hdrval{2}}] {};
  \node[leaf] (drop) at (7,2)
    [label=below:\small\pfalse{}] {};

  \path[below]
    (proto) edge node {} (ip1)
    (ip1) edge node {} (pt1)
    (ip2) edge node {} (pt2);
  \path[dashed,below]
    (proto) edge node {} (drop)
    (ip1) edge node {} (ip2)
    (ip2) edge node {} (drop);
\end{tikzpicture}
\caption{\proto{} $\sqsubset$ \textsf{dst}.}
\label{fdd-1}
\end{subfigure}
\quad
\begin{subfigure}[t]{0.2\textwidth}
\begin{tikzpicture}

  \node[node] (ip1) at (5,0)
    [label=right:\small\match{\dstip{}}{\hdrval{10.0.0.1}}] {};
  \node[node] (ip2) at (6,1)
    [label=right:\small\match{\dstip{}}{\hdrval{10.0.0.2}}] {};
  \node[node] (proto1) at (5,2)
    [label=right:\small\match{\proto{}}{\hdrval{http}}] {};
  \node[node] (proto2) at (6,2) {};

  \node[leaf] (pt1) at (5,3)
    [label=below:\small\modify{\ptf{}}{\hdrval{1}}] {};
  \node[leaf] (pt2) at (6,3)
    [label=below:\small\modify{\ptf{}}{\hdrval{2}}] {};
  \node[leaf] (drop) at (7,3)
    [label=below:\small\pfalse{}] {};

  \path[below]
    (ip1) edge node {} (proto1)
    (proto1) edge node {} (pt1)
    (ip2) edge node {} (proto2)
    (proto2) edge node{} (pt2);
  \path[dashed,below]
    (ip1) edge node {} (ip2)
    (proto1) edge node {} (drop)
    (ip2) edge node {} (drop)
    (proto2) edge node {} (drop);

\end{tikzpicture}
\caption{\textsf{dst} $\sqsubset$ \proto{}.}
\label{fdd-2}
\end{subfigure}

\caption{Two ordered \fdds for the same program.}
\end{figure}
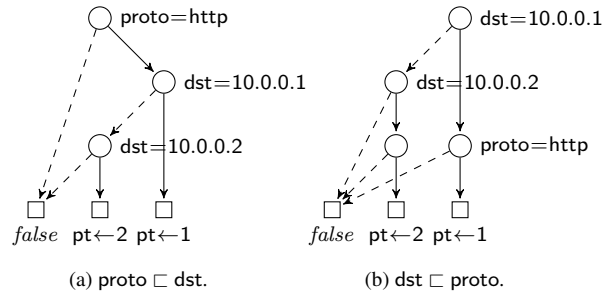

\begin{figure*}
\noindent
\center
\yellowbox{
\begin{minipage}{.5\textwidth}
\(
\begin{array}{@{}r@{~~}r@{~}c@{~}l@{\qquad}l@{}}
\multicolumn{4}{@{}l}{\textbf{Syntax}}\\[.5em]
\textrm{Booleans} & b & \Coloneqq  & \top \mid \bot\\[.125em]
\textrm{Contexts} & \Gamma & \Coloneqq & \cdot \mid \Gamma,(f,n):b\\[.125em]
\textrm{Actions} & a & \Coloneqq & \set{\modify{\field_1}{n_1}, \dots, \modify{\field_k}{n_k}} \\[.125em]
\textrm{Diagrams} & d & \Coloneqq & \set{a_1,\dots,a_k}               & \textit{Constant} \\
                  &   & \mid      & \ite{\match{\field}{n}}{d_1}{d_2} & \textit{Conditional}
\end{array}
\)
\medskip
\hrule
\medskip
\(
\begin{array}{@{}r@{\,}c@{\,}l@{}l@{}}
\multicolumn{4}{@{}l}{\textbf{Semantics}}\\[.5em]
\den{\set{\modify{\field_1}{n_1}, \dots, \modify{\field_k}{n_k}}}~(\hcons{\pk}{\h}) & \defeq & \set{\hcons{\pkupd{\pkupd{\pk}{\field_1}{n_1} \cdots}{\field_k}{n_k}}{\h}}\\[.5em]
\den{\set{a_1,\dots,a_k}}~(\hcons{\pk}{\h}) & \defeq & \den{a_1}~(\hcons{\pk}{\h}) \cup \dots \cup \den{a_k}~(\hcons{\pk}{\h})\\[.5em]
\den{\ite{\match{\field}{n}}{d_1}{d_2}}~(\hcons{\pk}{\h}) & \defeq &
\begin{cases}
\den{d_1}~(\hcons{\pk}{\h}) & \text{if}~\pkproj{\pk}{\field} = n\\
\den{d_2}~(\hcons{\pk}{\h}) & \text{otherwise}
\end{cases}
\end{array}
\)
\end{minipage}\quad\vrule\;\;\begin{minipage}{.45\textwidth}
\noindent\textbf{Well Formedness}
\begin{mathpar}
\fbox{$\Gamma \sqsubset (f,n)$}\hfill\vspace*{-3em}
\\
\inferrule*[Right=Nil]
{ \quad }
{ \cdot \sqsubset (\field,n) }
\\
\inferrule*[Right=Lt]
{ \field' \sqsubset \field }
{ \Gamma,(\field',n'):b' \sqsubset (\field,n) }

\inferrule*[Right=Eq]
{ \field' = \field \\ n' \sqsubset n }
{ \Gamma,(\field',n'):\bot \sqsubset (\field,n) }
\\
\fbox{$\Gamma \vdash d$}\hfill\vspace*{-3em}
\\
\inferrule*[Right=Constant]
{ \quad }
{ \Gamma \vdash \set{a_1,\dots,a_k} }
\\
\inferrule*[Right=Conditional]
{ 
 \Gamma \sqsubset (\field,n)\\\\
 \Gamma, (\field, n) : \top \vdash d_1\\\\
 \Gamma, (\field, n) : \bot \vdash d_2
}
{ \Gamma \vdash \ite{\match{\field}{n}}{d_1}{d_2} }
\end{mathpar}
\end{minipage}
}
\caption{Forwarding decision diagrams: syntax, semantics, and well formedness.}
\label{fig:fdd}
\end{figure*}

\paragraph*{Binary decision diagrams.}
To avoid these issues, our compiler is based on a novel representation
of packet-forwarding functions using a generalization of
\emph{binary decision diagrams}
(\bdds)~\cite{Akers:1978:BDD:1310167.1310815,Bryant:1986:GAB:6432.6433}. To
briefly review, a \bdd is a data structure that encodes a
boolean function as a directed acyclic graph. The interior nodes
encode boolean variables and have two outgoing edges: a true edge
drawn as a solid line, and a false edge drawn as a dashed line. The
leaf nodes encode constant values true or false. Given an assignment
to the variables, we can evaluate the expression by following the
appropriate edges in the graph. An \emph{ordered} \bdd imposes
a total order in which the variables are visited. In general, the
choice of variable-order can have a dramatic effect on the size of
a \bdd and hence on the run-time of \bdd-manipulating
operations. Picking an optimal variable-order is \textsc{np}-hard, but
efficient heuristics often work well in
practice. A \emph{reduced} \bdd has no isomorphic subgraphs
and every interior node has two distinct successors. A \bdd
can be reduced by repeatedly applying these two transformations:
\begin{itemize}
\item If two subgraphs are isomorphic, delete one by connecting its
  incoming edges to the isomorphic nodes in the other,
  thereby \emph{sharing} a single copy of the subgraph.
\item If both outgoing edges of an interior node lead to the same
  successor, eliminate the interior node by connecting its incoming edges
  directly to the common successor node.
\end{itemize}
Logically, an interior node can be thought of as representing
an \textsc{if-then-else} expression.\footnote{We write conditionals as
\ite{a}{b}{c}, in the style of the C ternary operator.}
For example, the expression:
\begin{align*}
\ite{a}{\ite{c}{1}{\ite{d}{1}{0}}}{\ite{b}{\ite{c}{1}{\ite{d}{1}{0}}}{0}}
\end{align*}
represents a \bdd for the boolean expression $(a \vee
b) \wedge (c \vee d)$. This notation makes the logical structure of
the \bdd clear while abstracting away from the sharing in the
underlying graph representation and is convenient for
defining \bdd-manipulating algorithms.

In principle, we could use \bdds to directly encode \netkat programs as
follows. We would treat packet headers as flat, $n$-bit vectors and
encode \netkat predicates as $n$-variable \bdds. Since \netkat
programs produce sets of packets, we could represent them in a
relational style using \bdds with $2n$ variables. However, there are
two issues with this representation:

\begin{itemize}
  \item Typical \netkat programs modify only a few headers and leave
  the rest unchanged. The \bdd that represents such a program would
  have to encode the identity relation between most of its
  input-output variables. Encoding the identity relation with \bdds
  requires a linear amount of space, so even trivial programs, such as
  the identity program, would require large \bdds.

  \item The final step of compilation needs to produce a prioritized
  flow table. It is not clear how to efficiently translate \bdds that
  represent \netkat programs as relations into tables that represent
  packet-processing functions. For example, a table of length one is
  sufficient to represent the identity program, but to generate this
  table from the \bdd sketched above, several paths would have to be
  compressed into a single rule.
\end{itemize}

\begin{figure*}
\smallskip
\noindent
\yellowbox{
\(
\begin{array}{@{}l@{}r@{\,}c@{\,}l@{}}
\text{\fbox{\(\lunion{d_1}{d_2}\)}}
& \lunion{\set{a_{11},\dots,a_{1k}}}{\set{a_{21},\dots,a_{2l}}} & \defeq &
 \set{a_{11},\dots,a_{1k}} \cup \set{a_{21},\dots,a_{2l}}\\[.25em]
& \lunion{\ite{\match{\field}{n}}{d_{11}}{d_{12}}}{\set{a_{21}, \dots a_{2l}}} & \defeq &
 \ite{\match{\field}{n}}{\lunion{d_{11}}{\set{a_{21}, \dots a_{2l}}}}{\lunion{d_{12}}{\set{a_{21}, \dots a_{2l}}}}\\[.25em]
& \lunion{\ite{\match{\field_1}{n_1}}{d_{11}}{d_{12}}}
       {\ite{\match{\field_2}{n_2}}{d_{21}}{d_{22}}} & \defeq &
\left\{
\begin{array}{l}
\ite{\match{\field_1}{n_1}}{\lunion{d_{11}}{d_{21}}}{\lunion{d_{12}}{d_{22}}} \qquad\qquad\qquad\quad\quad\,
  \text{if}~\field_1 = \field_2~\text{and}~n_1 = n_2\\[.125em]
\ite{\match{\field_1}{n_1}}{\lunion{d_{11}}{d_{22}}}{\lunion{d_{12}}{\ite{\match{\field_2}{n_2}}{d_{21}}{d_{22}}}} \quad\;\;\;
  \text{if}~\field_1 = \field_2~\text{and}~n_1 \sqsubset n_2\\[.125em]
\ite{\match{\field_1}{n_1}}{\lunion{d_{11}}{\ite{\match{\field_2}{n_2}}{d_{21}}{d_{22}}}}{\lunion{d_{12}}{\ite{\match{\field_2}{n_2}}{d_{21}}{d_{22}}}} \;\;\; 
  \text{if}~\field_1 \sqsubset \field_2\\
\end{array}\right.\\[.25em]
\multicolumn{4}{l}{\text{(omitting symmetric cases)}}\\[.25em]
\ \\[-.75em]
\hline
\ \\[-.75em]
\text{\fbox{\ltfilter{d}}}
& \ltfilter{\set{a_1,\dots,a_k}} & \defeq & \ite{\fnz}{\set{a_1,\dots,a_k}}{\set{}}\\[.25em]
& \ltfilter{\itei} & \defeq &
\begin{cases}
\ite{\fnz}{d_{11}}{\set{}} &
 \text{if}~\field=\field_1~\text{and}~n=n_1\\[.125em]
\ltfilter{(d_{12})} &
 \text{if}~\field=\field_1~\text{and}~n \neq n_1\\[.125em]
\ite{\fnz}{\itei}{\set{}} &
 \text{if}~\field \sqsubset \field_1\\[.125em]
\ite{\fni}{\ltfilter{(d_{11})}}{\ltfilter{(d_{12})}} &
 \text{otherwise}\\
\end{cases}\\
\ \\[-.75em]
\hline
\ \\[-.75em]
\multicolumn{4}{@{}l@{}}{
\begin{array}{@{}l@{~}|@{~}l@{}}
\begin{array}{@{}l@{~}r@{~}c@{~}l@{}}
\text{\fbox{\lseq{d_1}{d_2}}}
& \lseqmod{a}{\set{a_1,\dots,a_k}} &\defeq&
  \set{\lseq{a}{a_1},\dots,\lseq{a}{a_k}}\\[.25em]
& \lseqmod{a}{\ite{\match{\field}{n}}{d_1}{d_2}} &\defeq&
\begin{cases}
\lseqmod{a}{d_1} & \text{if}~\modify{\field}{n} \in a \\[.125em]
\lseqmod{a}{d_2} & \text{if}~\modify{\field}{n'} \in a \wedge n' \neq n\\[.125em]
\ite{\match{\field}{n}}{\lseqmod{a}{d_1}}{\lseqmod{a}{d_2}} & \text{otherwise}
\end{cases}\\[.25em]
& \lseq{\set{a_{1},\dots,a_{k}}}{d} & \defeq & \lunion{\lseqmod{a_{1}}{d}}{\dots \lunion{}{\lseqmod{a_{k}}{d}}}\\[.25em]
& \lseq{\ite{\match{\field}{n}}{d_{11}}{d_{12}}}{d_2} & \defeq & \lunion{\ltfilter{(\lseq{d_{11}}{d_2})}}{\lffilter{(\lseq{d_{12}}{d_2})}}\\
\end{array} &
\begin{array}{@{}l@{~}r@{~}c@{~}l@{}}
\text{\fbox{\lnegate{d}}}
& \lnegate{\set{}} & \defeq & \set{\set{}}\\[.25em]
& \lnegate{\set{a_1,\dots,a_k}} & \defeq & \set{}~\text{where}~k \geq 1\\[.25em]
& \lnegate{\itez} & \defeq & \ite{\fnz}{\lnegate{d_1}}{\lnegate{d_2}}\\
\ \\[-.75em]
\hline
\ \\[-.75em]
\text{\fbox{\lstar{d}}} & \lstar{d} & \defeq & \mathsf{fix}~(\lambda d'.\,\set{\set{}} + \lseq{d}{d'})\\[2.5em]
\end{array}
\end{array}
}
\end{array}
\)
}
\caption{Auxiliary definitions for local compilation to \fdds.}
\label{fig:fdd-helpers}
\end{figure*}

\paragraph*{Forwarding Decision Diagrams.}
To encode \netkat programs as decision diagrams, we introduce a modest
generalization of \bdds called \emph{forwarding decision diagrams}
(\fdds).  An \fdd differs from \bdds in two ways. First, interior
nodes match header fields instead of individual bits, which means we
need far fewer variables compared to a \bdd to represent the same
program. Our \fdd implementation requires $12$ variables (because
OpenFlow supports $12$ headers), but these headers span over $200$
bits. Second, leaf nodes in an \fdd directly encode packet
modifications instead of boolean values. Hence, \fdds do not encode
programs in a relational style.

Figures~\ref{fdd-1}~and~\ref{fdd-2} show \fdds for a program that
forwards \textsc{http} packets to hosts \hdrval{10.0.0.1}
and \hdrval{10.0.0.2} at ports \hdrval{1} and \hdrval{2}
respectively. The diagrams have interior nodes that match on headers
and leaf nodes corresponding to the actions used in the program.

To generalize ordered \bdds to \fdds, we assume orderings on fields
and values, both written $\sqsubset$, and lift them to tests
$\match{\field}{n}$ lexicographically:
\[
\fni \sqsubset \fnii \defeq (\field_1 \sqsubset \field_2) \vee (\field_1 = \field_2 \wedge n_1 \sqsubset n_2)
\]
We require that tests be arranged in ascending order from the
root. For reduced \fdds, we stipulate that they must have no
isomorphic subgraphs and that each interior node must have two unique
successors, as with \bdds, and we also require that
the \fdd must not contain redundant tests and
modifications. For example, if the
test \match{\dstip{}}{\hdrval{10.0.0.1}} is true,
then \match{\dstip{}}{\hdrval{10.0.0.2}} must be false. Accordingly,
an \fdd should not perform the latter test if the former
succeeds. Similarly, because \netkat's union operator ($\punion{p}{q}$)
is associative, commutative, and idempotent, to broadcast packets to
both ports 1 and 2 we could either
write \punion{\modify{\ptf{}}{\hdrval{1}}}{\modify{\ptf{}}{\hdrval{2}}}
or \punion{\modify{\ptf{}}{\hdrval{2}}}{\modify{\ptf{}}{\hdrval{1}}}. Likewise,
repeated modifications to the same header are equivalent to just the
final modification, and modifications to different headers
commute. Hence, updating the \dstip{} header to \hdrval{10.0.0.1} and
then immediately re-updating it to \hdrval{10.0.0.2} is the same as
updating it to \hdrval{10.0.0.2}.
In our implementation, we enforce the conditions for ordered,
reduced \fdds by representing actions as sets of sets of
modifications, and by using smart constructors that eliminate
isomorphic subgraphs and contradictory tests.

Figure~\ref{fig:fdd} summarizes the syntax, semantics, and
well-formedness conditions for \textsc{fdds} formally. Syntactically,
an \fdd $d$ is either a constant diagram specified by a set of actions
$\set{a_1,\dots,a_k}$, where an action $a$ is a finite map
$\{\modify{f_1}{n_1},\dots,\modify{f_k}{n_k}\}$ from fields to values
such that each field occurs at most once; or a conditional diagram
$\itez$ specified by a test $\fnz$ and two sub-diagrams. Semantically,
an action $a$ denotes a sequence of modifications, a constant diagram
$\set{a_1,\dots,a_k}$ denotes the union of the individual actions, and
a conditional diagram $\itez$ tests if the packet satisfies the
test and evaluates the true branch ($d_1$) or false branch ($d_2$)
accordingly. The well-formedness judgments $\Gamma \sqsubset (f,n)$
and $\Gamma \vdash d$ ensure that tests appear in ascending order and
do not contradict previous tests to the same field. The context
$\Gamma$ keeps track of previous tests and boolean outcomes.

\begin{figure}
\noindent
\yellowbox{
\begin{minipage}{.96\columnwidth}
\begin{alignat*}{2}
\compile{\pfalse}                   &\defeq \set{}&
\compile{\modify{\field}{n}}        &\defeq \set{\set{\modify{\field}{n}}}\\
\compile{\ptrue}                    &\defeq \set{\set{}}&
\compile{\match{\field}{n}}         &\defeq
  \ite{\match{\field}{n}}{\set{\set{}}}{\set{}}\\
\compile{\pnot{\polp}}              &\defeq {\lnegate{\compile{\polp}}}&
\compile{\punion{\polp_1}{\polp_2}} &\defeq
  \lunion{\compile{\polp_1}}{\compile{\polp_2}}\\
\compile{\pstar{\polp}}             &\defeq {\lstar{\compile{\polp}}} \qquad
  \quad&
\compile{\pseq{\polp_1}{\polp_2}}   &\defeq
  \lseq{\compile{\polp_1}}{\compile{\polp_2}}\\[-.75em]
\end{alignat*}
\end{minipage}
}
\caption{Local compilation to \fdds.}
\label{fig:localfdd}
\end{figure}

\paragraph*{Local compiler.}
Now we are ready to present the local compiler itself, which goes in
two stages. The first stage translates \netkat source programs
into \fdds, using the simple recursive translation given in
Figures~\ref{fig:fdd-helpers} and \ref{fig:localfdd}.

\noindent The \netkat primitives $\ptrue$, $\pfalse$, and $\modify{\field}{n}$
all compile to simple constant \fdds. Note that the empty action set
$\set{}$ drops all packets while the singleton action set
$\set{\set{}}$ containing the identity action $\set{}$ copies packets
verbatim. \netkat tests $\fnz$ compile to a conditional whose branches
are the constant diagrams for $\ptrue$ and $\pfalse$
respectively. \netkat union, sequence, negation, and star all
recursively compile their sub-programs and combine the results using
corresponding operations on \fdds, which are given in
Figure~\ref{fig:fdd-helpers}.

The \fdd union operator $(\lunion{d_1}{d_2})$ walks down the structure
of $d_1$ and $d_2$ and takes the union of the action sets at the
leaves. However, the definition is a bit involved as some care is
needed to preserve well-formedness. In particular, when combining
multiple conditional diagrams into one, one must ensure that the
ordering on tests is respected and that the final diagram does not
contain contradictions. Readers familiar with \bdds may notice that
this function is simply the standard ``apply'' operation (instantiated
with union at the leaves). The sequential composition operator
$(\lseq{d_1}{d_2})$ merges two packet-processing functions into a
single function. It uses auxiliary operations $\ltfilter{d}$ and
$\lffilter{d}$ to restrict a diagram $d$ by a positive or negative
test respectively. We elide the sequence operator on atomic actions
(which behaves like a right-biased merge of finite maps) and the
negative restriction operator (which is similar to positive
restriction, but not identical due to contradictory tests) to save
space. The first few cases of the sequence operator handle situations
where a single action on the left is composed with a diagram on the
right. When the diagram on the right is a conditional,
$\ite{\match{\field}{n}}{d_1}{d_2}$, we partially evaluate the test
using the modifications contained in the action on the left. For
example, if the left-action contains the
modification \modify{\field}{n}, we know that the test will be true,
whereas if the left-action modifies the field to another value, we
know the test will be false. The case that handles sequential
composition of a conditional diagram on the left is also
interesting. It uses restriction and union to implement the
composition, reordering and removing contradictory tests as needed to
ensure well formedness. The negation $\lnegate{d}$ operator is defined
in the obvious way. Note that because negation can only be applied to
predicates, the leaves of the diagram $d$ are either $\set{}$ or
$\set{\set{}}$. Finally, the \fdd Kleene star operator $\lstar{d}$ is
defined using a straightforward fixed-point computation. The
well-formedness conditions on \fdds ensures that a fixed point exists.

The soundness of local compilation from \netkat programs to \fdds is
captured by the following theorem:

\begin{theorem}[Local Soundness]
If $\compile{p} = d$ then $\den{p}~\h
= \den{d}~\h$.
\end{theorem}
\vspace*{-1.5\medskipamount}
\begin{proof} 
Straightforward induction on $p$.
\end{proof}

\begin{figure}
\begin{minipage}{0.36\columnwidth}
\begin{tikzpicture}

  \node[node] (proto) at (6,-1)
    [label=right:\small\match{\proto{}}{\hdrval{http}}] {};
  \node[node] (ip1) at (5,0)
    [label=right:\small\match{\dstip{}}{\hdrval{10.0.0.1}}] {};
  \node[node] (ip2) at (6,1)
    [label=right:\small\match{\dstip{}}{\hdrval{10.0.0.2}}] {};

  \node[leaf] (pt1) at (5,2)
    [label=below:\small\modify{\ptf{}}{\hdrval{1}}] {};
  \node[leaf] (pt2) at (6,2)
    [label=below:\small\modify{\ptf{}}{\hdrval{2}}] {};
  \node[leaf] (drop) at (7,2)
    [label=below:\small\pfalse{}] {};

  \path[below]
    (proto) edge node {} (ip1)
    (ip1) edge node {} (pt1)
    (ip2) edge node {} (pt2);
  \path[dashed,below]
    (proto) edge node {} (drop)
    (ip1) edge node {} (ip2)
    (ip2) edge node {} (drop);

\end{tikzpicture}
\end{minipage}
\quad
\begin{minipage}{0.45\columnwidth}
\vspace*{4em}
\[
\small
\begin{flowtable}
\match{\proto{}}{\hdrval{http}}, \match{\dstip{}}{\hdrval{10.0.0.1}} & \modify{\ptf{}}{\hdrval{1}} \\
\match{\proto{}}{\hdrval{http}}, \match{\dstip{}}{\hdrval{10.0.0.2}} & \modify{\ptf{}}{\hdrval{2}} \\
\match{\proto{}}{\hdrval{http}}  & \pfalse \\
\star                            & \pfalse
\end{flowtable}
\]
\end{minipage}
\caption{Forwarding table generation example.}\label{flowtablegen}
\end{figure}
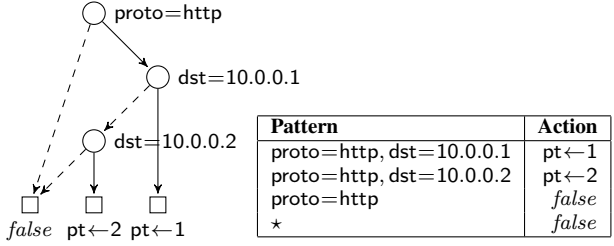

The second stage of local compilation converts \fdds to forwarding
tables. By design, this transformation is mostly straightforward: we
generate a forwarding rule for every path from the root to a leaf,
using the conjunction of tests along the path as the pattern and the
actions at the leaf. For example, the \fdd in
Figure~\ref{flowtablegen} has four paths from the root to the leaves
so the resulting forwarding table has four rules. The left-most path
is the highest-priority rule and the right-most path is the
lowest-priority rule. Traversing paths from left to right has the
effect of traversing true-branches before their associated
false-branches. This makes sense, since the only way to encode a
negative predicate is to partially shadow a negative-rule with a
positive-rule. For example, the last rule in the figure cannot encode
the test $\nmatch{\proto}{\hdrval{http}}$. However, since that rule is
preceded by a pattern that tests $\match{\proto}{\hdrval{http}}$, we
can reason that the \proto{} field is not \textsc{http} in the last
rule. If performed naively, this strategy could create a lot of extra
forwarding rules---e.g., the table in Figure~\ref{flowtablegen} has
two drop rules, even though one of them completely shadows the
other. In section~\ref{classbench}, we discuss optimizations that
eliminate redundant rules, exploiting the \fdd representation.

\section{Global Compilation\label{s:global}}




Thus far, we have seen how to compile local \netkat programs into
forwarding tables using \fdds. Now we turn to the global compiler,
which translates global programs into equivalent local programs.

In general, the translation from global to local programs requires
introducing extra state, since global programs may use regular
expressions to describe end-to-end forwarding paths---e.g., recall the
example of a global program with two overlapping paths from
Section~\ref{sec:overview}. Put another way, because a local program
does not contain \pdup, the compiler can analyze the entire program
and generate an equivalent forwarding table that executes on a single
switch, whereas the control flow of a global program must be made
explicit so execution can be distributed across multiple switches.
More formally, a local program encodes a function from packets
to sets of packets, whereas a global program encodes a function
from packets to sets of packet-histories.

To generate the extra state needed to encode the control flow of a
global, distributed execution into a local program, the global
compiler translates programs into finite state automata. To a first
approximation, the automaton can be thought of as the one
for the regular expression embedded in the global program, and the
instrumented local program can be thought of as encoding the states
and transitions of that automaton in a special header field. The actual
construction is a bit more complex for several reasons. First, we
cannot instrument the topology in the same way that we instrument
switch terms. Second, we have to be careful not to introduce extra
states that may lead to duplicate packet histories being generated.
Third, \netkat programs have more structure than ordinary regular
expressions, since they denote functions on packet histories rather
than sets of strings, so a more complicated notion of automaton---a symbolic \netkat automaton---is
needed.

At a high-level, the global compiler proceeds in several steps:
\begin{itemize}

\item It compiles the input program to an equivalent symbolic 
  automaton. All valid paths through
  the automaton alternate between switch-processing states and
  topology-processing states, which enables executing them as local
  programs.

\item It introduces a \emph{program counter} by instrumenting  the 
automaton to keep track of the current automaton state in
the \textit{pc} field.

\item It determinizes the \netkat automaton using an analogue of
the subset construction for finite automata.

\item It uses heuristic optimizations to reduce the number of states.

\item It merges all switch-processing states into a single switch
state and all topology-processing states into a single topology state.
\end{itemize}
The final result is a single local program that can be compiled using
the local compiler. This program is equivalent to the original global
program, modulo the \textit{pc} field, which records the automaton
state.

\subsection{\netkat Automata}
In prior work, some of the authors introduced \netkat automata and
proved the analogue of Kleene's theorem: programs and automata have
the same expressive power~\cite{netkat-automata}.  This allows us to
use automata as an intermediate representation for arbitrary
\netkat programs. This section reviews \netkat automata, which 
are used in the global compiler, and then presents a function that
constructs an automaton from an arbitrary \netkat program.

\begin{definition}[\netkat Automaton]
A \netkat automaton is a tuple $(S,s_0,\epsilon,\delta)$, where:
\begin{itemize}
\item $S$ is
a finite set of states, 
\item $s_0 \in S$ is the start state, 
\item $\epsilon :
S \rightarrow \Pk \rightarrow \powerset{\Pk}$ is the
\emph{observation function}, and
\item $\delta : S \rightarrow \Pk \rightarrow \powerset{\Pk \times S}$ is
the \emph{continuation function}.
\end{itemize}
\end{definition}
\noindent A \netkat automaton is 
said to be \emph{deterministic} if $\delta$ maps each packet to a
unique next state at every state, or more formally if
\[
|\set{s' : S \mid (pk',s') \in \delta\, s\, pk}| \leq 1
\]
for all states $s$ and packets $pk$ and $pk'$.

The inputs to \netkat automata are guarded strings drawn from the set
$\Pseq{\Pk, \pstar{(\pseq{\Pk}{\pdup})}, \Pk}$. That is, the inputs
have the form
\[
  \Pseq{\pk_{\mathit{in}},\pk_1,\pdup,\pk_2,\pdup,\cdot,\pk_n,\pdup,\pk_{\mathit{out}}}
\]
where $n \geq 0$. Intuitively, such strings represent packet-histories
through a network: $\pk_{\mathit{in}}$ is the input state of a packet,
$\pk_{\mathit{out}}$ is the output state, and the $\pk_{i}$ are the intermediate
states of the packet that are recorded as it travels through the network. 

To process such a string, an automaton in state $s$ can
either \emph{accept} the trace if $n=0$ and $\pk_{\mathit{out}} \in \epsilon\,
s\, \pk_{\mathit{in}}$, or it can consume one packet and $\pdup$ from the start
of the string and transition to state $s'$ if $n>0$ and
$(pk_1,s') \in \delta\, s\, \pk_{\mathit{in}}$. In the latter case, the
automaton yields a residual trace:
\[
  \Pseq{\pk_{1},\pk_2,\pdup,\cdot,\pk_n,\pdup,\pk_{\mathit{out}}}
\]
Note that the ``output'' $\pk_1$ of state $s$ becomes the ``input''
to the successor state $s'$.
More formally, acceptance is defined as:
\begin{align*}
  \accept s\, (\Pseq{\pk_{\mathit{in}},\pk_{\mathit{out}}}) &\quad \Leftrightarrow \quad
    \pk_{\mathit{out}} \in \epsilon\, s\, \pk_{\mathit{in}}\\
  \accept s\, (\Pseq{\pk_{\mathit{in}},\pk_1,\pdup,w}) &\quad \Leftrightarrow \quad
    \bigvee_{\mathclap{(\pk_1,s')\, \in\, \delta\, s\, \pk_{\mathit{in}}}}
    \accept s' (\Pseq{\pk_1,w})
\end{align*}

Next, we define a function that builds an automaton $\auto{\polp}$
from an arbitrary \netkat program $\polp$ such that
\begin{multline*}
  (\pk_{\mathit{out}} \cons \pk_n \cons \dots \cons \hone{\pk_1})
  \in \den{p} \hone{pk_{\mathit{in}}}\\
  \Leftrightarrow \quad
  \accept_{A(p)} s_0\, (\Pseq{\pk_{\mathit{in}},\pk_1,\pdup,\dots,\pk_{\mathit{out}}})
\end{multline*}
The construction is based on Antimirov partial derivatives for regular
expressions \cite{antimirov:partial}.  We fix a set of labels $L$, and
annotate each occurrence of $\pdup$ in the source program $\polp$ with
a unique label $\ell \in L$. We then define a pair of functions:
\begin{itemize}
\item $\globale{\cdot} : \mathsf{Pol} \rightarrow \mathsf{Pol}$ and 
\item $\globald{\cdot} : \mathsf{Pol} \rightarrow \powerset{\mathsf{Pol}\times L\times \mathsf{Pol}}$ 
\end{itemize}
Intuitively, $\globale{\polp}$ can be thought of as extracting the
local components from $p$ (and will be used to construct $\epsilon$),
while $\globald{\polp}$ extracts the global components (and will be
used to construct $\delta$). A triple
$\triple{d}{\ell}{k} \in \globald{\polp}$ represents the derivative of
$p$ with respect to $\pdup^{\ell}$. That is, $d$ is the \pdup-free
component of $p$ up to $\pdup^{\ell}$, and $k$ is the residual program
(or \emph{continuation}) of $p$ after $\pdup^{\ell}$.

We calculate $\globale{\polp}$ and $\globald{\polp}$ simultaneously
using a simple recursive algorithm defined in
Figure~\ref{fig:automata}. The definition makes use of the following
abbreviations,
\begin{align*}
  \pseq{\globald{\polp}}{\polq} &\defeq
    \set{\triple{d}{\ell}{\pseq{k}{\polq}} \mid
         \triple{d}{\ell}{k} \in \globald{\polp}}\\
  \pseq{\polq}{\globald{\polp}} &\defeq
    \set{\triple{\pseq{\polq}{d}}{\ell}{k} \mid
         \triple{d}{\ell}{k} \in \globald{\polp}}
\end{align*}
which lift sequencing to sets of triples in the
obvious way.

\begin{figure}
\noindent
\centerline{
\yellowbox{
\(
\begin{array}{c|c|c}
\polp                 &\globale{\polp} : \mathsf{Pol}
                      &\globald{\polp} : \powerset{\mathsf{Pol} \times L \times \mathsf{Pol}}\\\hline
a                     &a                   &\emptyset  \\
\modify{\field}{n}    &\modify{\field}{n}  &\emptyset  \\
\pdup^{\ell}          &\pfalse             &\set{\triple{\ptrue}{\ell}{\ptrue}}\\
\punion{\polq}{\polr} &\punion{\globale{\polq}}{\globale{\polr}}
                      &\globald{\polq} \cup {\globald{\polr}} \\
\pseq{\polq}{\polr}   &\pseq{\globale{\polq}}{\globale{\polr}}
                      &\pseq{\globald{\polq}}{r}
                       \cup \pseq{\globale{\polq}}{\globald{\polr}}\\
\pstar{\polq}         &\globale{\polq}*
                      &\Pseq{
                          \globale{\pstar{\polq}},
                          \globald{\polq},
                          \pstar{\polq}
                       }
\end{array}
\)}}
\caption{Auxiliary definitions for \netkat automata construction.}
\label{fig:automata}
\end{figure}

The next lemma characterizes $\globale{\polp}$ and
$\globald{\polp}$, using the following notation to reconstruct
programs from sets of triples:

\[
\sum\globald{p} \defeq \sum_{\triple{d}{\ell}{k} \in \globald {p}} \Pseq{d, \pdup, k}
\]

\begin{lemma}[Characterization of $\globale{\cdot}$ and $\globald{\cdot}$]
\label{lem:e-d-characterization}
  For all programs $p$, we have the following:
  \begin{enumerate}[(a)]
    \item $p \equiv \punion{\globale{p}}{\sum\globald {p}}$.
    \item $\globale{p}$ is a local program.
    \item For all $\triple{d}{\ell}{k} \in \globald{p}$, $d$ is a local program.
    \item For all labels $\ell$ in $\polp$, there exist unique programs $d$ and
    $k$ such that $\triple{d}{\ell}{k} \in \globald{\polp}$.
  \end{enumerate}
\end{lemma}
\vspace*{-1.5\medskipamount}
\begin{proof}
By structural induction on $p$. Claims $(b-d)$ are trivial. Claim
$(a)$ can be proved purely equationally using only the \netkat axioms
and the \textsc{kat-denesting} rule from \cite{anderson:netkat}.
\end{proof}

\noindent Lemma~\ref{lem:e-d-characterization}~(d) allows us to write
$k_{\ell}$ to refer to the unique continuation of $\pdup^{\ell}$. By
convention, we let $k_{0}$ denote the ``initial continuation,'' namely
$\polp$.

\begin{definition}[Program Automaton]
The \netkat automaton $\auto{\polp}$ for a program $\polp$ is defined
as $(S,s_0,\epsilon,\delta)$ where
\begin{itemize}
\item $S$ is the set of labels occurring in $\polp$, plus the initial label $0$.
\item $s_0 \defeq 0$
\item $\epsilon\,\ell\,\pk \defeq
  \set{\pk' \mid \hone{pk'} \in \den{\globale{k_{\ell}}} \hone{pk}}$
\item $\delta\,\ell\,\pk \defeq
  \set{(\pk',\ell') \mid
          \triple{d}{\ell'}{k} \in \globald{k_{\ell}} \wedge
          \hone{pk'} \in \den{d} \hone{pk}}$
\end{itemize}
\end{definition}

\begin{theorem}[Program Automaton Soundness]
 For all programs $p$, packets $\pk$ and histories $\h$, we
 have \[ \h \in \den{p} \hone{\pk_{\mathit{in}}} \Leftrightarrow \accept s_0\,
 (\Pseq{\pk_{\mathit{in}}, \pk_1, \pdup, \cdots, \pk_n, \pdup, \pk_{\mathit{out}}})
\] where $h = pk_{\mathit{out}} \cons pk_{n} \cons \cdots \cons \hone{\pk_1}$.
\end{theorem}
\vspace*{-1.5\medskipamount}
\begin{proof}
We first strengthen the claim, replacing $\hone{pk_{\mathit{in}}}$ with an
arbitrary history $pk_{\mathit{in}} \cons \h'$, $s_0$ with an arbitrary label
$\ell \in S$, and $p$ with $k_{\ell}$. We then proceed by induction on the
length of the history, using {Lemma \ref{lem:e-d-characterization}}
for the base case and induction step.
\end{proof}

\subsection{Local Program Generation}

With a \netkat automaton $\auto{\polp}$ for the global program $\polp$
in hand, we are now ready to construct a local program. The main idea
is to make the state of the global automaton explicit in the local
program by introducing a new header field $\mathit{pc}$ (represented
concretely using \textsc{vlan}s, \textsc{mpls} tags, or any other
unused header field) that keeps track of the state as the packet traverses
the network. This encoding enables simulating the automaton for the
global program using a single local program (along with the physical
topology). We also discuss determinization and optimization, which are
important for correctness and performance.

\paragraph*{Program counter.}
The first step in local program generation is to encode the state of
the automaton into its observation and transition functions using the
$\mathit{pc}$ field. To do this, we use the same structures as are
used by the local compiler, \fdds. Recall that the observation
function $\epsilon$ maps input packets to output packets according to
$\mathcal{E}\den{k_{\ell}}$, which is a $\pdup$-free \netkat program. Hence,
we can encode the observation function for a given state $\ell$ as a
conditional \fdd that tests whether $\mathit{pc}$ is $\ell$ and either
behaves like the \fdd for $\mathcal{E}\den{k_{\ell}}$ or $\pfalse$. We can
encode the continuation function $\delta$ as an \fdd in a similar
fashion, although we also have to set the $\mathit{pc}$ to each
successor state $s'$.
This symbolic representation of automata
using \fdds allows us to efficiently manipulate automata despite
the large size of their ``input alphabet'', namely $|\Pk \times \Pk|$.
In our implementation we introduce the
$\mathit{pc}$ field and \fdds on the fly as automata are constructed,
rather than adding them as a post-processing step, as is described
here for ease of exposition.

\paragraph*{Determinization.}
The next step in local program generation is to determinize
the \netkat automaton. This step turns out to be critical for
correctness---it eliminates extra outputs that would be produced if we
attempted to directly implement a nondeterministic \netkat
automaton. To see why, consider a program of the form $p +
p$. Intuitively, because union is an idempotent operation, we 
expect that this program will behave the same as just a single copy
of $p$. However, this will not be the case when $p$ contains a
$\pdup$: each occurrence of $\pdup$ will be annotated with a different
label.  Therefore, when we instrument the program to track automaton
states, it will create two packets that are identical expect for the
$\mathit{pc}$ field, instead of one packet as required by the
semantics.  The solution to this problem is simply to determinize the
automaton before converting it to a local program. Determinization
ensures that every packet trace induces a unique path through the
automaton and prevents duplicate packets from being produced.
Using \fdds to represent the automaton symbolically is crucial for
this step: it allows us to implement a \netkat analogue of the subset
construction efficiently.

\paragraph*{Optimization.}
One practical issue with building automata using the algorithms
described so far is that they can use a large number of states---one
for each occurrence of \pdup{} in the program---and determinization
can increase the number of states by an exponential factor. Although
these automata are not wrong, attempting to compile them can lead to
practical problems since extra states will trigger a proliferation of
forwarding rules that must be installed on switches. Because switches
today often have limited amounts of memory---often only a few thousand
forwarding rules---reducing the number of states is an important
optimization. An obvious idea is to optimize the automaton using
(generalizations of) textbook minimization
algorithms. Unfortunately this would be prohibitively expensive since
deciding whether two states are equal is a costly operation in the
case of \netkat automata. Instead, we adopt a simple heuristic that
works well in practice and simply merge states that are identical. In
particular, by representing the observation and transition functions
as \fdds, which are hash consed, testing equality is cheap---simple
pointer comparisons.

\paragraph*{Local Program Extraction.}
The final step is to extract a local program from the automaton.
Recall from Section~\ref{sec:overview} that, by definition, links are
enclosed by $\pdup$s on either side, and links are the only
\netkat terms that contain $\pdup$s or modify the switch field.
It follows that every global program gives rise to a bipartite \netkat
automaton in which all accepting paths alternate between ``switch
states'' (which do not modify the switch field) and ``link states''
(which forward across links and do modify the switch field), beginning
with a switch state.  Intuitively, the local program we want to
extract is simply the union of of the $\epsilon$ and $\delta$ \fdds of
all switch states (recall Lemma~\ref{lem:e-d-characterization}~(a)), with
the link states implemented by the physical network. Note however,
that the physical network will neither match on the $pc$ nor advance
the $pc$ to the next state (while the link states in our automaton
do).  To fix the latter, we observe that any link state has a unique
successor state.  We can thus simply advance the $pc$ by two states
instead of one at every switch state, anticipating the missing $pc$
modification in link states.
To address the former, we employ the equivalence
\[
  \plink{\sw_1}{\pt_1}{\sw_2}{\pt_2} \equiv
  \Pseq{\match{\sw}{1},\match{\pt}{1}, t, \match{\sw}{2},\match{\pt}{2}}
\]
It allows us to replace links with the entire topology if we modify switch
states to match on the appropriate source and destination
locations immediately before and after transitioning across a link.
After modifying the $\epsilon$ and $\delta$ \fdds
accordingly and taking the union of all switch states as described
above, the resulting \fdd can be passed to the local compiler to
generate forwarding tables.

The tables will correctly implement the global program provided the physical
topology $(in,t,out)$ satisfies the following:
\begin{itemize}
  \item $p \equiv \Pseq{in,p,out}$, i.e. the global program specifies
  end-to-end forwarding paths
  \item $t$ implements at least the links used in $p$.
  \item $\pseq{t}{in} \equiv \pfalse \equiv \pseq{out}{t}$, \ie
  the $in$ and $out$ predicates should not include locations that are internal
  to the network.
\end{itemize}

\newcommand{\node}[2]{\ensuremath{\begin{bmatrix}#1\\#2\end{bmatrix}}}
\newcommand{\din}{\ensuremath{\mathtt{I}}\xspace}
\newcommand{\dout}{\ensuremath{\mathtt{O}}\xspace}
\newcommand{\step}{\ensuremath{\rightarrow}\xspace}
\newcommand{\vstep}{\ensuremath{\rightarrow_v}\xspace}
\newcommand{\pstep}{\ensuremath{\rightarrow_p}\xspace}
\newcommand{\vplayer}{\ensuremath{\mathcal{V}}\xspace}
\newcommand{\fplayer}{\ensuremath{\mathcal{F}}\xspace}
\newcommand{\floop}[1]{\ensuremath{\mathsf{Loop}\ #1}\xspace}
\newcommand{\fout}{\ensuremath{f_{\mathit{out}}}\xspace}
\newcommand{\fin}{\ensuremath{f_{\mathit{in}}}\xspace}

\section{Virtual Compilation}

The third and final stage of our compiler pipeline translates virtual
programs to physical programs. Recall that a virtual program is one
that is defined over a virtual topology. Network virtualization can
make programs easier to write by abstracting complex physical
topologies to simpler topologies and also makes programs portable
across different physical topologies. It can even be used to multiplex
several virtual networks onto a single physical network---\eg, in
multi-tenant datacenters~\cite{vmware:nvp}.

To compile a virtual program, the compiler needs to know the mapping
between virtual switches, ports, and links and their counterparts at
the physical level. The programmer supplies a virtual program
$\mathit{v}$, a virtual topology $\mathit{t}$, sets of ingress and
egress locations for $\mathit{t}$, and a relation $\mathcal{R}$
between virtual and physical ports. The relation $\mathcal{R}$ must
map each physical ingress to a virtual ingress, and conversely for
egresses, but is otherwise unconstrained---e.g., it need not be
injective or even a function.\footnote{Actually, we can relax this
condition slightly and allow physical ingresses to map to zero or one
virtual ingresses---if a physical ingress has no corresponding
representative in the virtual network, then packets arriving at that
ingress will not be admitted to the virtual network.} The constraints
on ingresses and egresses ensures that each packet entering the
physical network lifts uniquely to a packet in the virtual network,
and similarly for packets editing the virtual network. During
execution of the virtual program, each packet can be thought of as
having two locations, one in the virtual network and one in the
physical network; $\mathcal{R}$ defines which pairs of locations are
consistent with each other. For simplicity, we assume the virtual
program is a local program. If it is not, the programmer can use the
global compiler to put it into local form.

\paragraph*{Overview.}
To execute a virtual program on a physical network, possibly with a
different underlying topology, the compiler must (i) instrument the
program to keep track of packet locations in the virtual topology and
(ii) implement forwarding between locations that are adjacent in the
virtual topology using physical paths. To achieve this, the virtual
compiler proceeds as follows:
\begin{enumerate}
\item It instruments the program to use the virtual switch
 (\vswf) and virtual port (\vptf) fields that track of the location of
 the packet in the virtual topology.
\item It constructs a \emph{fabric}: a \netkat program that updates
the physical location of a packet when its virtual location changes
and vice versa, after each step of processing to restore consistency
with respect to the virtual-physical relation, $\mathcal{R}$.
\item It assembles the final program by combining $v$ with the fabric, 
eliminating the \vswf and \vptf fields, and compiling the result using
the global compiler.
\end{enumerate}
Most of the complexity arises in the second step because there may be
many valid fabrics (or there may be none).  However, this step is
independent of the virtual program. The fabric can be computed once
and for all and then be reused as the program changes. Fabrics can be
generated in several ways---\eg, to minimize a costs such as path
length or latency, maximize disjointness, etc.

\begin{figure}[t]
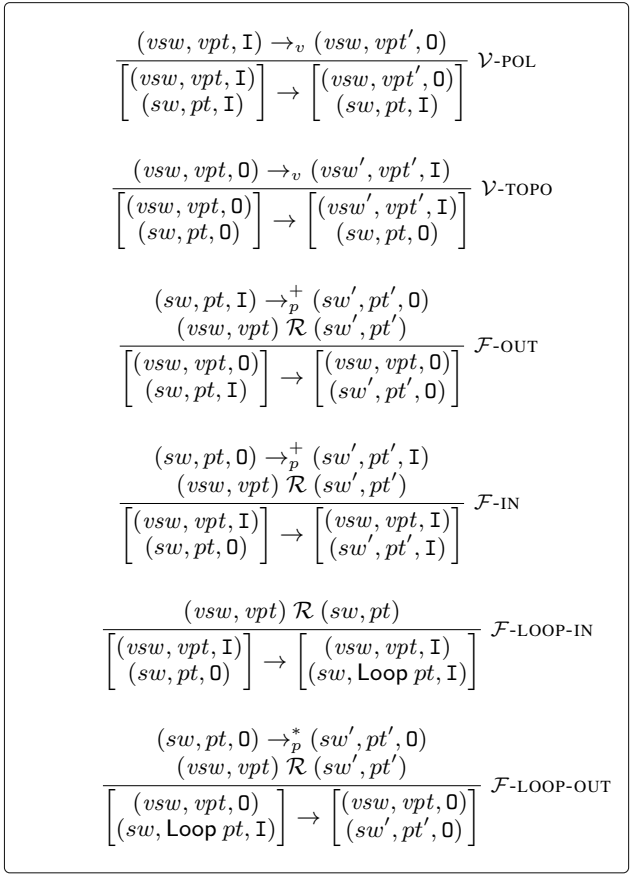

\noindent
\center
\yellowbox{
\begin{minipage}{.95\columnwidth}
\medskip
\qquad\;\;{\(
\begin{array}{@{\quad}c@{\quad}}
\inferrule*[Right=\vplayer-pol]
  {(\mathit{vsw},\mathit{vpt},\din) \vstep (\mathit{vsw},\mathit{vpt}',\dout)}
  {\node{(\mathit{vsw},\mathit{vpt},\din)}{(sw,pt,\din)} \step
   \node{(\mathit{vsw},\mathit{vpt}',\dout)}{(sw,pt,\din)}} \\[1.5em]
\inferrule*[Right=\vplayer-topo]
  {(\mathit{vsw},\mathit{vpt},\dout) \vstep (\mathit{vsw}',\mathit{vpt}',\din)}
  {\node{(\mathit{vsw},\mathit{vpt},\dout)}{(sw,pt,\dout)} \step
   \node{(\mathit{vsw}',\mathit{vpt}',\din)}{(sw,pt,\dout)}} \\[1.5em]
\inferrule*[Right=\fplayer-out]
  {(sw,pt,\din) \pstep^+ (sw',pt',\dout) \\\\ (\mathit{vsw},\mathit{vpt})~\mathcal{R}~(sw',pt')}
  {\node{(\mathit{vsw},\mathit{vpt},\dout)}{(sw,pt,\din)} \step
   \node{(\mathit{vsw},\mathit{vpt},\dout)}{(sw',pt',\dout)}} \\[1.5em]
\inferrule*[Right=\fplayer-in]
  {(sw,pt,\dout) \pstep^+ (sw',pt',\din) \\\\ (\mathit{vsw},\mathit{vpt})~\mathcal{R}~(sw',pt')}
  {\node{(\mathit{vsw},\mathit{vpt},\din)}{(sw,pt,\dout)} \step
   \node{(\mathit{vsw},\mathit{vpt},\din)}{(sw',pt',\din)}} \\[1.5em]
\inferrule*[Right=\fplayer-loop-in]
  {(\mathit{vsw},\mathit{vpt})~\mathcal{R}~(sw,pt)}
  {\node{(\mathit{vsw},\mathit{vpt},\din)}{(sw,pt,\dout)} \step
   \node{(\mathit{vsw},\mathit{vpt},\din)}{(sw,\floop{pt},\din)}} \\[1.5em]
\inferrule*[Right=\fplayer-loop-out]
  {(sw,pt,\dout) \pstep^* (sw',pt',\dout) \\\\ (\mathit{vsw},\mathit{vpt})~\mathcal{R}~(sw',pt')}
  {\node{(\mathit{vsw},\mathit{vpt},\dout)}{(sw,\floop{pt},\din)} \step
   \node{(\mathit{vsw},\mathit{vpt},\dout)}{(sw',pt',\dout)}}
\end{array}
\)}
\medskip
\end{minipage}
}
\caption{Fabric game graph edges.}
\label{edges}
\end{figure}

\begin{figure}[t]
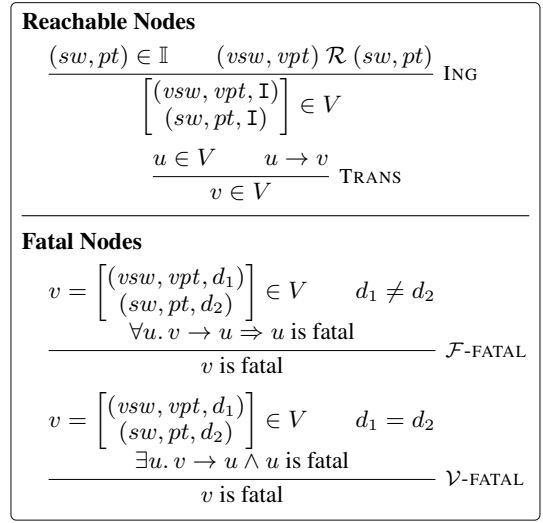

\noindent
\center
\yellowbox{
\begin{minipage}{.8\columnwidth}
\noindent{\textbf{Reachable Nodes}}\\[.5em]
\(
\begin{array}{@{\quad}c@{\quad}}
\inferrule*[Right=Ing]
  {(sw,pt) \in \mathbb{I} \\ (\mathit{vsw},\mathit{vpt})~\mathcal{R}~(sw,pt)}
  {\node{(\mathit{vsw},\mathit{vpt},\din)}{(sw,pt,\din)} \in V}
\\[.5em]
\inferrule*[Right=Trans]
  {u \in V \\ u \step v}
  {v \in V}
\\[.25em]
\end{array}
\)\\
\hrule
\ \\[.25em]
\noindent{\textbf{Fatal Nodes}}\\[.5em]
\(
\begin{array}{@{\quad}c@{\quad}}
\inferrule*[Right=\fplayer-fatal]
  {v=\node{(\mathit{vsw},\mathit{vpt},d_1)}{(sw,pt,d_2)} \in V \\ d_1 \neq d_2 \\\\
   \forall u.\, v \step u \implies u \text{ is fatal}}
  {v \text{ is fatal}}
\\[.5em]
\inferrule*[Right=\vplayer-fatal]
  {v=\node{(\mathit{vsw},\mathit{vpt},d_1)}{(sw,pt,d_2)} \in V \\ d_1 = d_2 \\\\
   \exists u.\, v \step u \land u \text{ is fatal}}
  {v \text{ is fatal}}
\end{array}
\)
\end{minipage}
}
\caption{Reachable and fatal nodes.}
\label{nodes}
\end{figure}

\paragraph*{Instrumentation.}
To keep track of a packet's location in the virtual network, we
introduce new packet fields \vswf and \vptf for the virtual switch and
the virtual port, respectively. We replace all occurrences of the \swf
or \ptf field in the program $v$ and the virtual topology $t$
with \vswf and \vptf respectively using a simple textual
substitution. Packets entering the physical network must be lifted to
the virtual network. Hence, we replace $\mathit{in}$ with a program
that matches on all physical ingress locations $\mathbb{I}$ and
initializes \vswf and \vptf in accordance with $\mathcal{R}$:
\[
    \mathit{in'} \defeq
    \sum_{\mathclap{\substack{(\mathit{sw},\mathit{pt}) \in \mathbb{I}\\
         (\mathit{vsw},\mathit{vpt})~\mathcal{R}~(\mathit{sw},\mathit{pt})}}}
        \Pseq{
            \match{\swf}{\mathit{sw}}, \match{\ptf}{\mathit{pt}},
            \modify{\vswf}{\mathit{vsw}}, \modify{\vptf}{\mathit{vpt}}
        }
\]
Recall that we require $\mathcal{R}$ to relate each location in
$\mathbb{I}$ to at most one virtual ingress, so the program lifts each
packet to at most one ingress location in the virtual network.
The \vswf and \vptf fields are only used to track locations during the
early stages of virtual compilation. They are completely eliminated in
the final assembly. Hence, we will not need to introduce additional
tags to implement the resulting physical program.

\paragraph*{Fabric construction.}
Each packet can be thought of as having two locations: one in the
virtual topology and one in the underlying physical topology. After
executing $\mathit{in'}$, the locations are consistent according to the
virtual-physical relation $\mathcal{R}$. However, consistency can be
broken after each step of processing using the virtual program $v$ or
virtual topology $t$. To restore consistency, we construct
a \emph{fabric} comprising programs \fin and \fout from the virtual
and physical topologies and $\mathcal{R}$, and insert it into the
program:
\[
q \defeq \Pseq{\mathit{in'}, (\pseq{v}{\fout}), \pstar{(\Pseq{t, \fin, v,
\fout})},
\mathit{out}}
\]
In this program, $v$ and $t$ alternate with \fout and \fin in
processing packets, thereby breaking and restoring consistency
repeatedly.  Intuitively, it is the job of the fabric to keep the
virtual and physical locations in sync.

This process can be viewed as a two-player game between a virtual
player \vplayer (embodied by $v$ and $t$) and a fabric player \fplayer
(embodied by \fout and \fin). The players take turns moving a packet
across the virtual and the physical topology, respectively.
Player \vplayer wins if the fabric player \fplayer fails to restore
consistency after a finite number of steps; player \fplayer wins
otherwise. Constructing a fabric now amounts to finding a winning
strategy for \fplayer.

We start by building the game graph $G=(V,E)$ modeling all possible
ways that consistency can be broken by \vplayer or restored by \fplayer.
Nodes are pairs of virtual and physical locations, $[l_v,l_p]$,
where a location is a 3-tuple comprising a switch, a port, and 
a direction that indicates if the packet entering the port (\din)
leaving the port (\dout). The rules in Figure~\ref{edges} determine the
edges of the game graph:
\begin{itemize}

  \item The edge $[l_v,l_p] \step [l_v',l_p]$ exists if \vplayer can
  move packets from $l_v$ to $l_v'$. There are two ways to do so:
  either \vplayer moves packets across a virtual switch
  (\textsc{\vplayer-pol}) or across a virtual link
  (\textsc{\vplayer-topo}). In the inference rules,  we write
  $\vstep$ to denote a single hop in the virtual topology:
  \[
  (\mathit{vsw}, \mathit{vpt}, d) \vstep (\mathit{vsw}',\mathit{vpt}',d')
  \]
  if $d=\din$ and $d'=\dout$ then the hop is across one switch, but if
   $d=\dout$ and $d'=\din$ then the hop is across a link.

  \item The edge $[l_v,l_p] \step [l_v,l_p']$ exists if \fplayer can move
  packets from $l_p$ to $l_p'$. When \fplayer makes a move, it must restore
  physical-virtual consistency (the $\mathcal{R}$ relation in the premise of
  \textsc{\fplayer-pol} and \textsc{\fplayer-topo}). To do so, it may
  need to take several
  hops through the physical network (written as $\pstep^+$).

  \item In addition, \fplayer may leave a packet at their
  current location, if the location is already consistent
  (\textsc{\fplayer-loop-in} and \textsc{\fplayer-loop-out}). Note
  that these force a packet located at physical location
  $(sw,pt,\dout)$ to leave through port $pt$ eventually. Intuitively,
  once the fabric has committed to emitting the packet through a given
  port, it can only delay but not withdraw that commitment.

\end{itemize}

Although these rules determine the complete game graph, all packets enter
the network at an ingress location (determined by the $\mathit{in}'$ predicate).
Therefore, we can restrict our attention to only those nodes that are
reachable from the ingress (reachable nodes in Figure~\ref{nodes}). 
In the resulting graph $G=(V,E)$, every path represents a possible
trajectory that a packet processed by $q$ may take through the virtual
and physical topology.

In addition to removing unreachable nodes, we must remove
\emph{fatal nodes}, which are the nodes where
\fplayer is unable to restore consistency and thus loses the game.
$\RefTirName{\fplayer-fatal}$ says that any state from which \fplayer
is unable to move to a non-fatal state is fatal. In particular, this
includes states in which \fplayer cannot move to any other state at
all.  $\RefTirName{\vplayer-fatal}$ says that any state in
which \vplayer can move to a fatal state is fatal.
Intuitively, we define such states to be fatal since we want the
fabric to work for any virtual program the programmer may write.
Fatal states can be removed using a simple backwards
traversal of the graph starting from nodes without outgoing
edges. This process may remove ingress nodes if they turn out to be
fatal. This happens if and only if there exists no fabric that can
always restore consistency for arbitrary virtual programs.  Of course,
this case can only arise if the physical topology is not
bidirectional.

\begin{figure*}
\begin{subfigure}{.33\textwidth}
\begin{tikzpicture}
\node{\pgfimage[width=\textwidth]{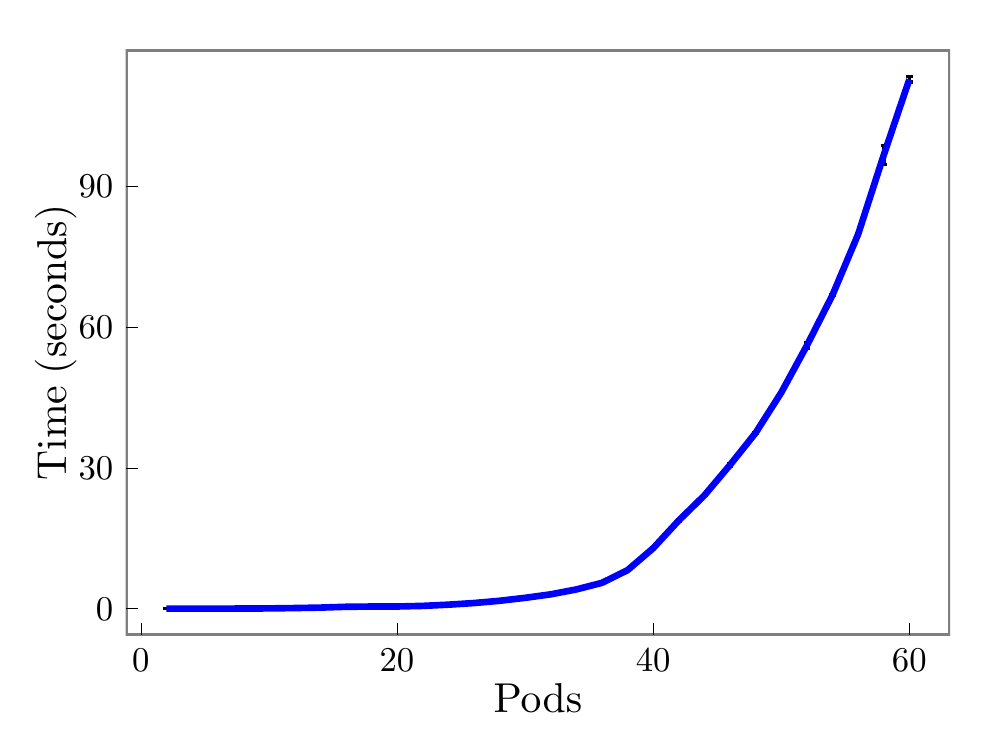}};
\end{tikzpicture}
\caption{Routing on $k$-pod fat-trees.}
\label{fattree_timing}
\end{subfigure}
\begin{subfigure}{.33\textwidth}
\begin{tikzpicture}
\node{\pgfimage[width=\textwidth]{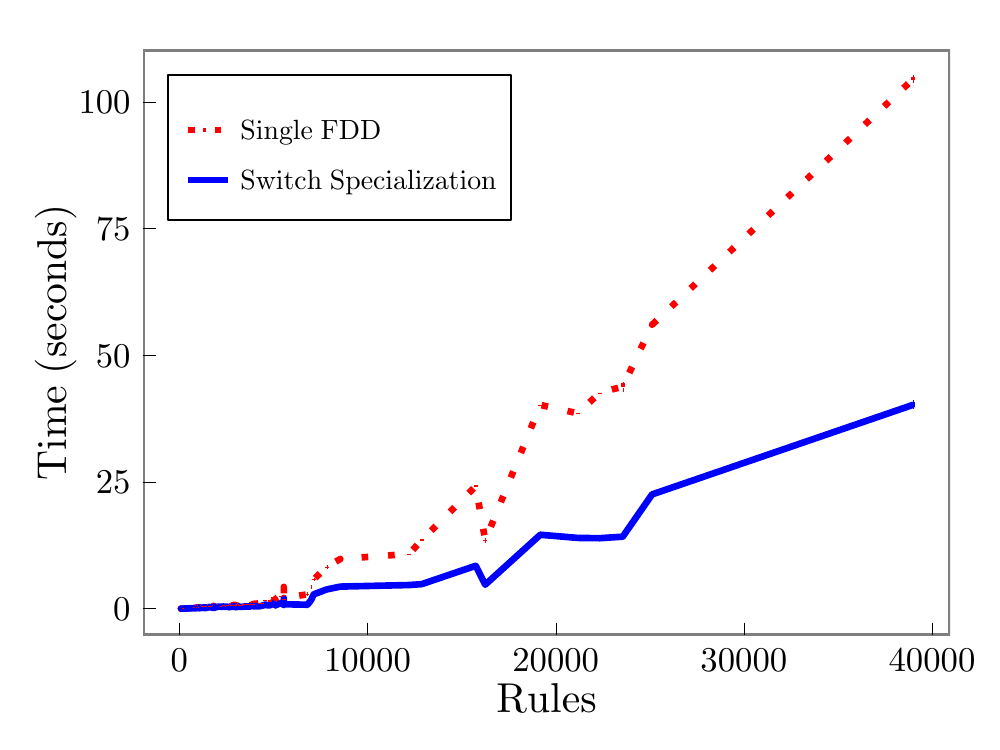}};
\end{tikzpicture}
\caption{Destination-based routing on topology zoo.}
\label{zoo-routing}
\end{subfigure}
\begin{subfigure}{.33\textwidth}
\begin{tikzpicture}
\node{\pgfimage[width=\textwidth]{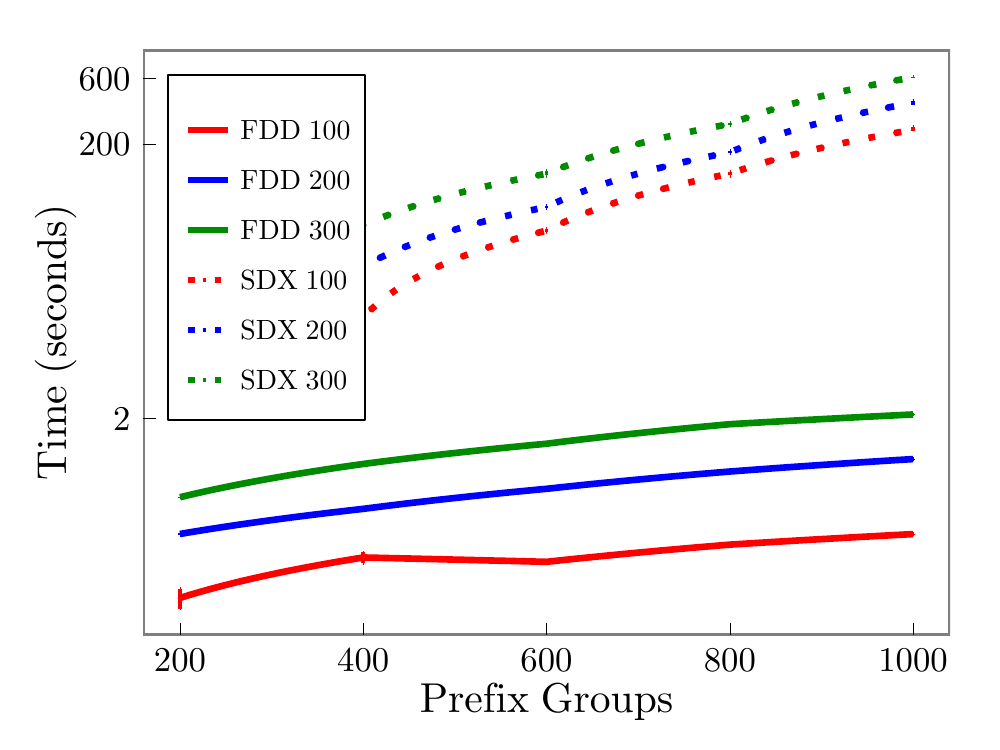}};
\end{tikzpicture}
\caption{Time needed to compile \textsc{sdx} benchmarks.}
\label{sdxtiming}
\end{subfigure}
\caption{Experimental results: compilation time.}
\end{figure*}

\paragraph*{Fabric selection.}

If all ingress nodes withstand pruning, the resulting graph encodes exactly the
set of all winning strategies for \fplayer, \ie the set of all possible fabrics.
A fabric is a subgraph of $G$ that contains the ingress, is closed under
all possible moves by the virtual program, and contains exactly one edge out of
every state in which \fplayer has to restore consistency. The \fplayer-edges
must be labeled with concrete paths through the physical topology, as there may
exist several paths implementing the necessary multi-step transportation from
the source node to the target node.

In general, there may be many fabrics possible and the choice of different \fplayer-edges
correspond to fabrics with different characteristics, such as minimizing hop counts,
maximizing disjoint paths, and so on. Our compiler implements several simple strategies. 
For example, given a metric $\phi$ on paths (such as hop count), our 
greedy strategy starts at the ingresses and adds a node
whenever it is reachable through an edge $e$ rooted at a node $u$
already selected, and e is (i) any \vplayer-player edge or (ii) the
\fplayer-player edge with path $\pi$ minimizing $\phi$ among all edges
and their paths rooted at $u$.

After a fabric is selected, it is straightforward to encode it as
a \netkat term. Every \fplayer-edge $[l_v,l_p] \step [l_v, l_p']$ in
the graph is encoded as a \netkat term that matches on the locations
$l_v$ and $l_p$, forwards along the corresponding physical path from
$l_p$ to $l_p'$, and then resets the virtual location to
$l_v$. Resetting the virtual location is semantically redundant but
will make it easy to eliminating the \vswf and \vptf fields.  We then
take $f_{\mathit{in}}$ to be the union of all
$\RefTirName{\fplayer-in}$-edges, and $f_{\mathit{out}}$ to be the
union of all $\RefTirName{\fplayer-out}$-edges. \netkat's global
abstractions play a key role, providing the building blocks for
composing multiple overlapping paths into a unified fabric.



\paragraph*{End-to-end Compilation.}
After the programs $\mathit{in'}$, $\fin$, and $\fout$, are calculated
from $\mathcal{R}$, we assemble the physical program $q$, defined
above. However, one last potential problem remains: although the virtual
compiler adds instrumentation to update the physical switch and port
fields, the program still matches and updates the virtual switch
(\vswf) and virtual port (\vptf). However, note that by construction
of $q$, any match on the \vswf or \vptf field is preceded by a
modification of those fields on the same physical switch. Therefore,
all matches are automatically eliminated during \fdd generation, and
only modifications of the $\vswf$ and $\vptf$ fields remain. These can
be safely erased before generating flow tables as the global compiler
inserts a program counter into $q$ that plays double-duty to track
both the physical location and the virtual location of a
packet. Hence, we only need a single tag to compile virtual programs!

\begin{figure*}
\begin{subfigure}{.33\textwidth}
\begin{tikzpicture}
\node{\pgfimage[width=\textwidth]{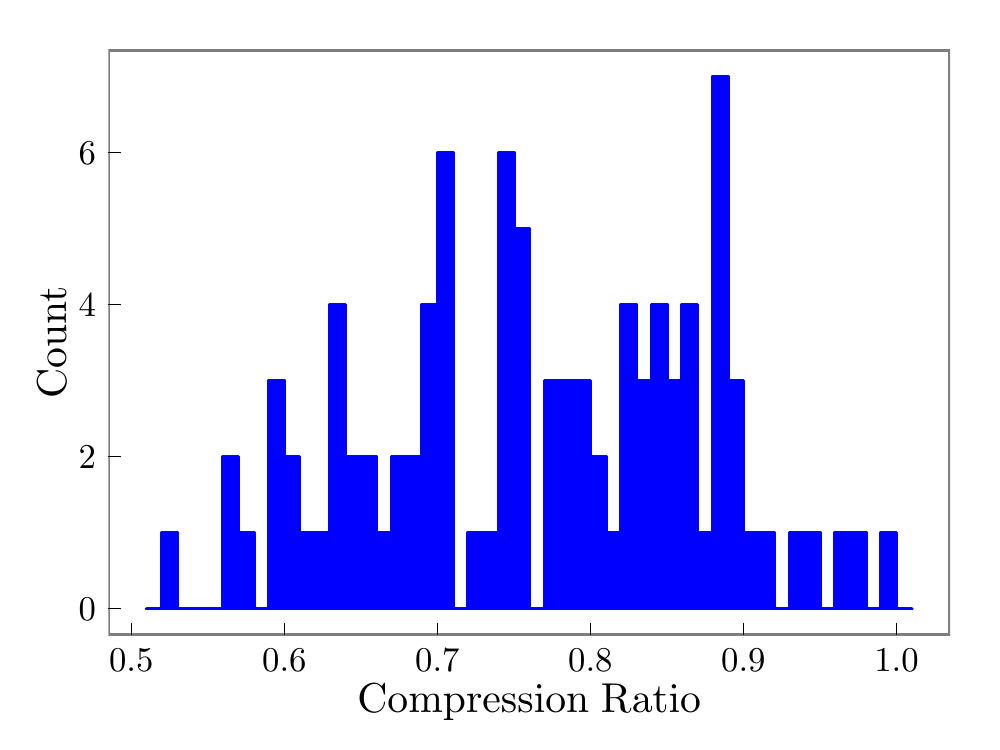}};
\end{tikzpicture}
\caption{Compressing Classbench \textsc{acl}s.}
\label{compression}
\end{subfigure}
\begin{subfigure}{.33\textwidth}
\begin{tikzpicture}
\node{\pgfimage[width=\textwidth]{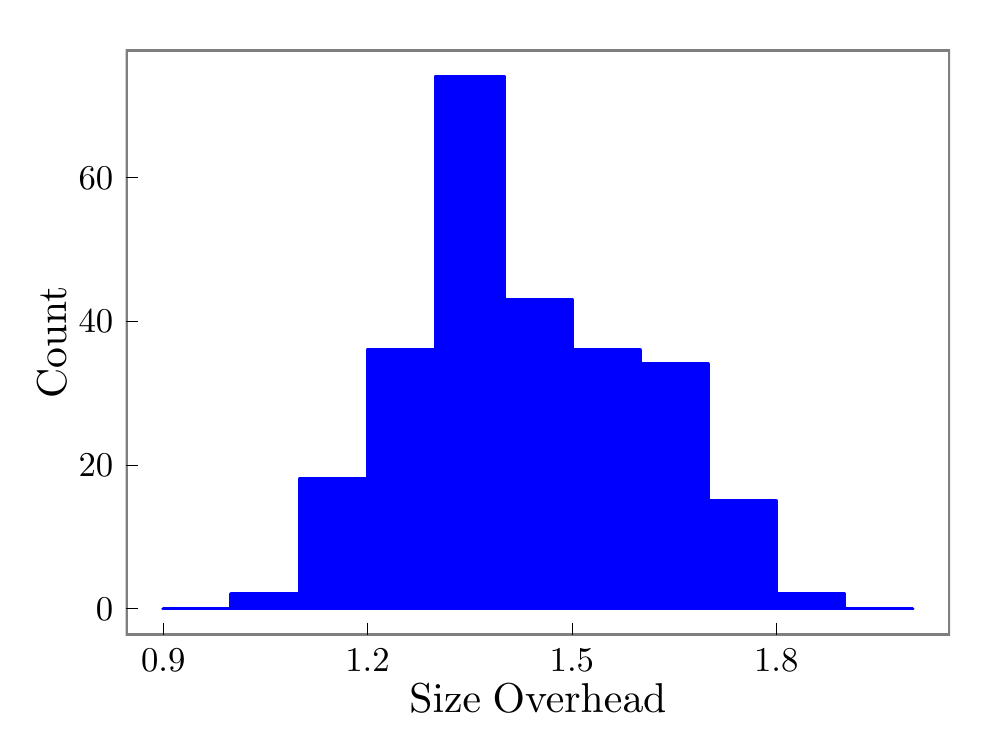}};
\end{tikzpicture}
\caption{Table size overhead for global programs.}\label{global_size_overhead}
\end{subfigure}
\begin{subfigure}{.33\textwidth}
\begin{tikzpicture}
\node{\pgfimage[width=\textwidth]{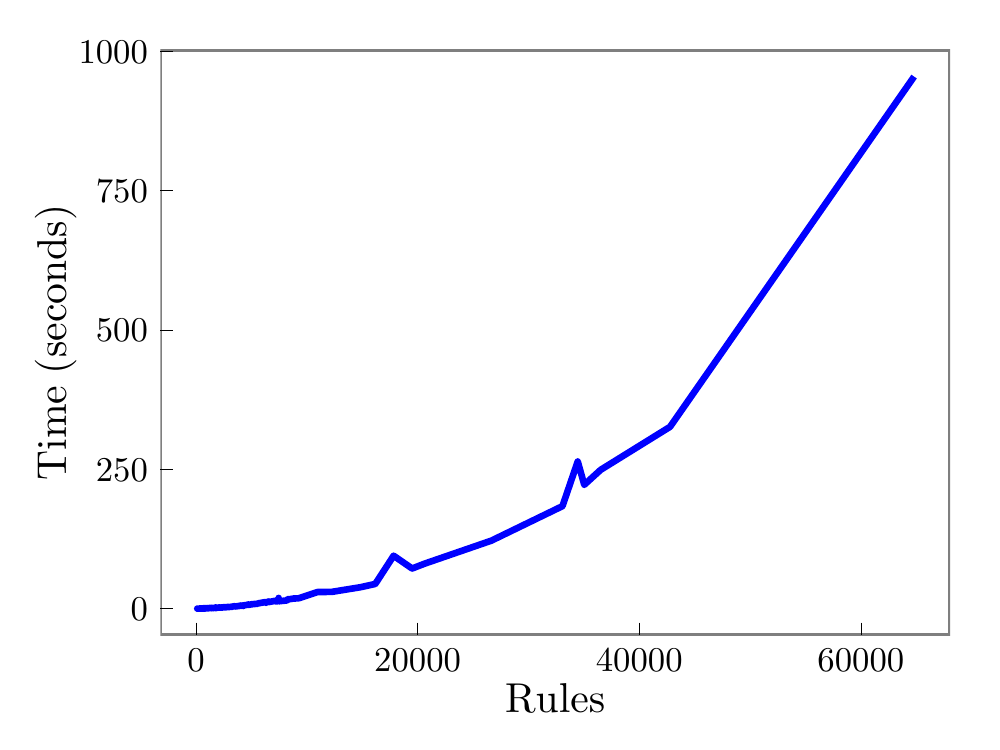}};
\end{tikzpicture}
\caption{Compilation time for global programs.}\label{global_time}
\end{subfigure}
\caption{Experimental results: forwarding table compression and global compilation.}
\end{figure*}

\section{Evaluation\label{eval}}

To evaluate our compiler, we conducted experiments on a diverse set of
real-world topologies and benchmarks. In practice, our compiler is a
module that is used by the Frenetic \textsc{sdn} controller to map \netkat
programs to flow tables. Whenever network events occur, \eg, a host
connects, a link fails, traffic patterns change, and so on, the
controller may react by generating a new \netkat program. Since network
events may occur rapidly, a slow compiler can easily be a bottleneck
that prevents the controller from reacting quickly to network events.
In addition, the flow tables that the compiler generates must be small
enough to fit on the available switches. Moreover, as small tables can
be updated faster than large tables, table size affects the
controller's reaction time too.

Therefore, in all the following experiments we measure flow-table
compilation time and flow-table size. We apply the compiler to
programs for a variety of topologies, from topology designs for very
large datacenters to a dataset of real-world topologies. We highlight
the effect of important optimizations to the fundamental \fdd-based
algorithms. We perform all experiments on 32-core, 2.6 GHz Intel Xeon
E5-2650 machines with 64GB RAM.\footnote{Our compiler is
single-threaded and doesn't leverage multicore.}  We repeat all timing
experiments ten times and plot their average.

\paragraph*{Fat trees.}
A fat-tree~\cite{fattree} is a modern datacenter network design that
uses commodity switches to minimize cost. It provides several
redundant paths between hosts that can be used to maximize available
bandwidth, provide backup paths, and so on. A fat-tree is organized
into pods, where a $k$-pod fat-tree topology can support up to
$\frac{k^3}{4}$ hosts. A real-world datacenter might have up to 48
pods~\cite{fattree}. Therefore, our compiler should be able to
generate forwarding programs for a 48-pod fat tree relatively quickly.

Figure~\ref{fattree_timing} shows how the time needed to generate all
flow tables varies with the number of pods in a
fat-tree.\footnote{This benchmark uses the switch-specialization
optimization, which we describe in the next section.} The graph shows
that we take approximately 30 seconds to produce tables for 48-pod fat
trees (\ie, 27,000 hosts) and less than 120 seconds to generate
programs for 60-pod fat trees (\ie, 54,000 hosts).

This experiment shows that the compiler can generate tables for large
datacenters. But, this is partly because the fat-tree forwarding
algorithm is topology-dependent and leverages symmetries to minimize
the amount of forwarding rules needed. Many real-world topologies are
not regular and require topology-independent forwarding programs. In
the next section, we demonstrate that our compiler scales well with
these topologies too.

\paragraph*{Topology Zoo.\label{zoo}}
The \emph{Topology Zoo}~\cite{topozoo} is a dataset of a few hundred
real-world network topologies of varying size and structure. For every
topology in this dataset, we use \emph{destination-based routing} to
connect all nodes to each other. In destination-based routing, each
switch filters packets by their destination address and forwards them
along a spanning-tree rooted at the destination. Since each switch
must be able to forward to any destination, the total number of rules
must be $\mathcal{O}(n^2)$ for an $n$-node network.

Figure~\ref{zoo-routing} shows how the running time of the compiler
varies across the topology zoo benchmarks. The curves are not as
smooth as the curve for fat-trees, since the complexity of forwarding
depends on features of network topology. Since the topology zoo is so
diverse, this is a good suite to exercise the \emph{switch
specialization} optimization that dramatically reduces compile time.

A direct implementation builds of the local compiler builds one \fdd
for the entire network and uses it to generate flow tables for each switch.
However, since several \fdd (and \bdd) algorithms are
fundamentally quadratic, it helps to first specialize the program for each
switch and then generate a small \fdd for each switch in the network
(\emph{switch specialization}). Building \fdds for several
smaller programs is typically much faster than building a single \fdd
for the entire network. As the graph shows, this optimization
has a dramatic effect on all but the smallest topologies.

\paragraph*{SDX.}
Our experiments thus far have considered some quite large forwarding
programs, but none of them leverage software-defined networking in any
interesting way.  In this section, we report on our performance on
benchmarks from a recent \textsc{sigcomm} paper~\cite{gupta:sdx} that
proposes a new application of \textsc{sdn}.

An Internet exchange point (\textsc{ixp}) is a physical location where networks
from several \textsc{isp}s connect to each other to exchange traffic.
Legal contracts between networks are often implemented by routing programs at
\textsc{ixp}s. However, today's \textsc{ixp}s use baroque protocols the needlessly
limit the kinds of programs that can be implemented.  A
 Software-defined \textsc{ixp} (an ``\textsc{sdx}''~\cite{gupta:sdx})
 gives participants fine-grained control over packet-processing and
 peering using a high-level network programming
 language. The \textsc{sdx} prototype uses
 Pyretic~\cite{monsanto:pyretic} to encode policies and presents
 several examples that demonstrate the power of an expressive network
 programming language.

We build a translator from Pyretic to \netkat and use it to evaluate
our compiler on \textsc{sdx}s own benchmarks. These benchmarks
simulate a large \textsc{ixp} where a few hundred peers apply programs
to several hundred prefix groups. The dashed lines in
Figure~\ref{sdxtiming} reproduce a graph from the
\textsc{sdx} paper, which shows how compilation time varies with the number
of prefix groups and the number of participants in the \textsc{sdx}.\footnote{We
get nearly the same numbers as the \textsc{sdx} paper on our hardware.} The
solid lines show that our compiler is orders of magnitude faster. Pyretic takes
over 10 minutes to compile the largest benchmark, but our compiler only takes
two seconds.

Although Pyretic is written in Python, which is a lot slower than
OCaml, the main problem is that Pyretic has a simple table-based
compiler that does not scale (Section~\ref{tablecompiler}). In fact,
the authors of \textsc{sdx} had to add several optimizations to get
the graph depicted. Despite these optimizations, our \fdd-based
approach is substantially faster.

The \textsc{sdx} paper also reports flow-table sizes for the same
benchmark. At first, our compiler appeared to produce tables that were
twice as large as Pyretic. Naturally, we were unhappy with this result
and investigated. Our investigation revealed a bug in the Pyretic
compiler, which would produce incorrect tables that were artificially
small. The authors of \textsc{sdx} have confirmed this bug and it has
been fixed in later versions of Pyretic. We are actively working with
them to port \textsc{sdx} to \netkat to help \textsc{sdx} scale
further.

\begin{figure*}
\begin{subfigure}{.33\textwidth}
\includegraphics[width=\textwidth
  ]{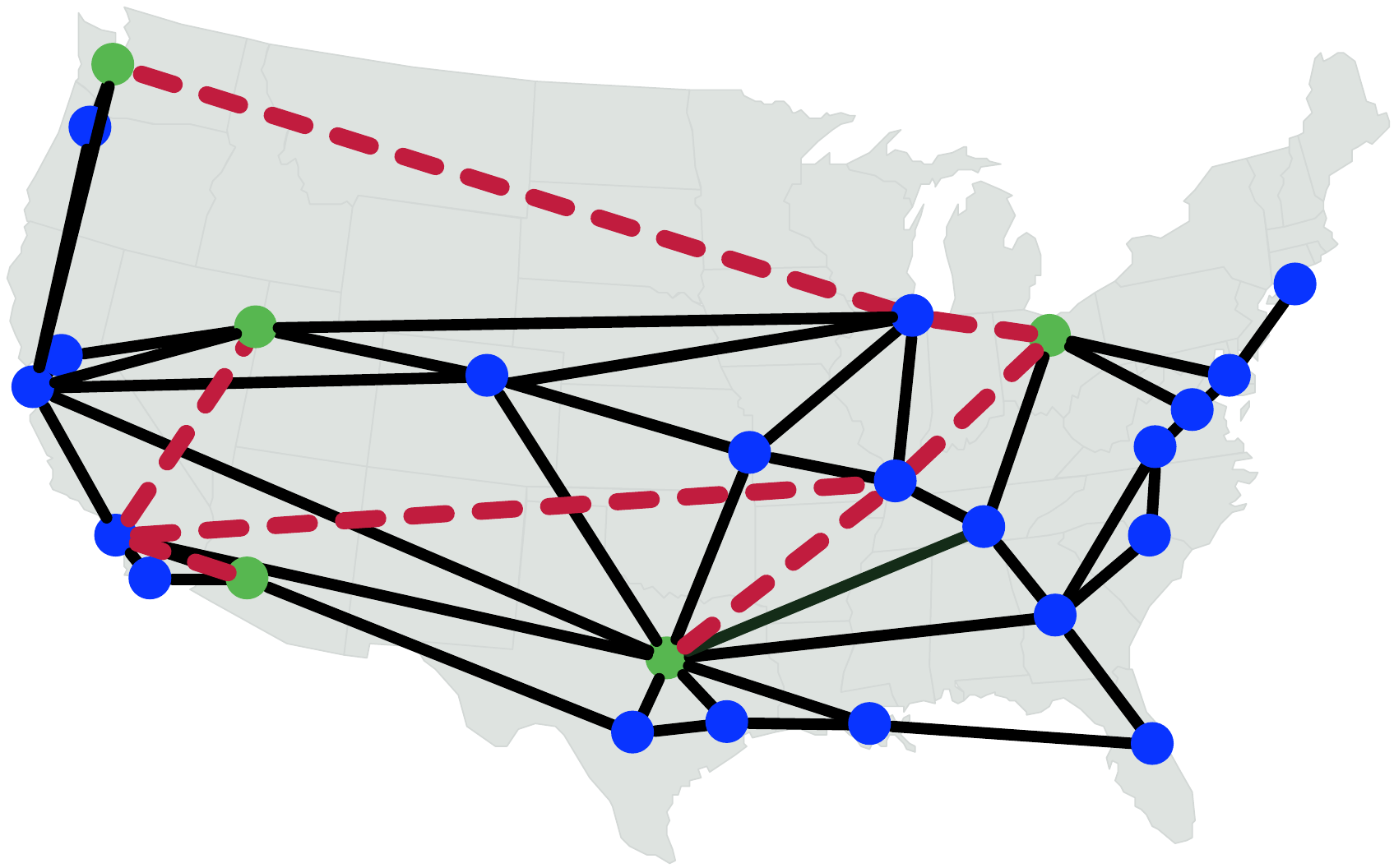}
\caption{minimum total number of links}
\end{subfigure}
\begin{subfigure}{.33\textwidth}
\includegraphics[width=\textwidth
  ]{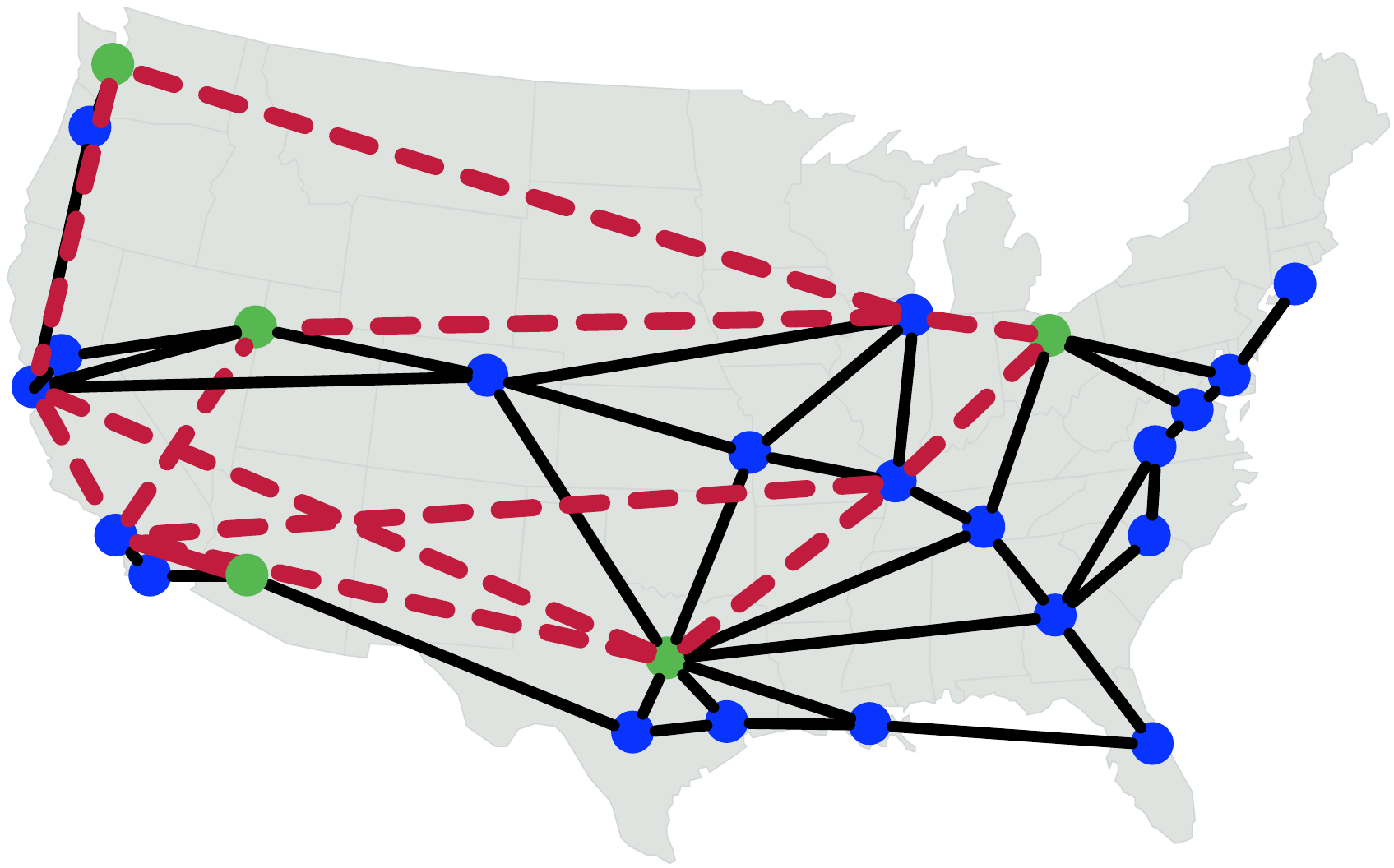}
\caption{minimum number of hops}
\end{subfigure}
\begin{subfigure}{.33\textwidth}
\includegraphics[width=\textwidth
  ]{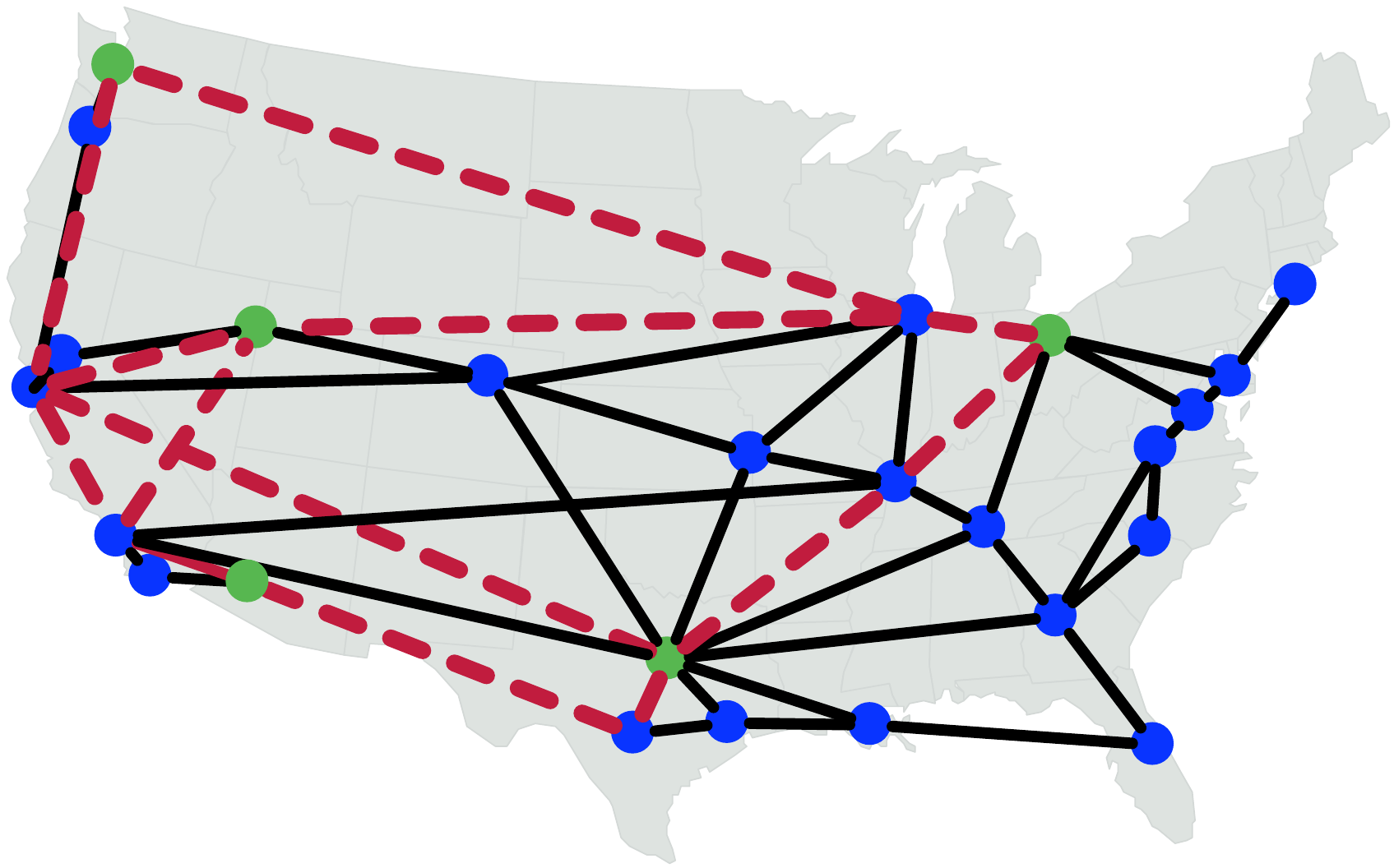}
\caption{minimum distance}
\end{subfigure}
\caption{Three fabrics optimizing different metrics\label{att}}
\end{figure*}

\paragraph*{Classbench\label{classbench}.}

Lastly, we compile \textsc{acl}s generated
using \emph{Classbench}~\cite{taylor:classbench}. These are realistic
firewall rules that showcase another optimization: it is often
possible to significantly compress tables by combining and eliminating
redundant rules.

We build an optimizer for the flow-table generation algorithm in
Figure~\ref{flowtablegen}.  Recall that that we generate flow-tables
by converting every complete path in the \fdd into a rule. Once a path
has been traversed, we can remove it from the \fdd without
harm. However, naively removing a path may produce an \fdd that is not
reduced. Our optimization is simple: we remove paths from the \fdd as
they are turned into rules and ensure that the \fdd is reduced at each
step. When the last path is turned into a rule, we are left with a
trivial \fdd. This iterative procedure prevents several unnecessary
rules from being generated. It is possible to implement other
canonical optimizations. But, this optimization is unique because it
leverages properties of reduced \fdds. Figure~\ref{compression} shows
that this approach can produce 30\% fewer rules on average than a
direct implementation of flow-table generation. We do not report
running times for the optimizer, but it is negligible in all our
experiments.

\paragraph*{Global compiler.}
The benchmarks discussed so far only use the local compiler. In this
section, we focus on the global compiler. Since the global compiler
introduces new abstractions, we can't apply it to existing benchmarks,
such as \textsc{sdx}, which use local programs. Instead, we need to
build our own benchmark suite of global programs. To do so, we build a
generator that produces global programs that describe paths between
hosts. Again, an $n$-node topology has $O(n^2)$ paths.  We apply this
generator to the Topology Zoo, measuring compilation time and table
size:

\begin{itemize}

\item \emph{Compilation time:} since the global compiler leverages \fdds, we
can expect automaton generation to be fast. However, global
compilation involves other steps such as determinization and
localization and their effects on compilation time may
matter. Figure~\ref{global_time} shows how compilation time varies
with the total number of rules generated.  This graph does grow faster
than local compilation time on the same benchmark (the red, dashed
line in Figure~\ref{zoo-routing}).  Switch-specialization, which
dramatically reduces the size of \fdds and hence compilation time,
does not work on global programs. Therefore, it makes most sense to
compare this graph to local compilation with a single \fdd.

\item \emph{Table size:} The global compiler has some
optimizations to eliminate unnecessary states, which produces fewer
rules.  However, it it does not fully minimize \netkat automata thus
it may produce more rules than equivalent local
programs. Figure~\ref{global_size_overhead} shows that on the topology
zoo, global routing produces tables that are no more than twice as
large as local routing.

\end{itemize}

\noindent We belive these results are promising: we spent a lot of time
tuning the local compiler, but the global compiler is an early
prototype with much room for improvement.

\paragraph*{Virtualization case study.}
Finally, we present a small case study that showcases the virtual
compiler on a snapshot of the AT\&T backbone network circa
2007--2008. This network is part of the Topology Zoo and shown in
Figure~\ref{att}. We construct a ``one big switch'' virtual network
and use it to connect five nodes (highlighted in green) to each other:
\[ 
\sum_{n=1}^{5} \pseq{\match{\dstf}{10.0.0.n}}{\modify{\ptf}{n}}
\]
To map the virtual network to the physical network, we generate three
different fabrics: (a) a fabric that minimizes the total number of
links used across the network, (b) a fabric that minimizes the number
of hops between hosts, and (c) a fabric that minimizes the physical
length of the path between hosts. In the figure, the links utilized by
each of these fabrics is highlighted in red.

The three fabrics give rise to three very different implementations of
the same virtual program. Note that the program and the fabric are
completely independent of each other and can be updated
independently. For example, the operator managing the physical network
could change the fabric to implement a new
\textsc{sla}, \eg move from minimum-utilization to shortest-paths. This
change requires no update to the virtual program; the network would
witness performance improvement for free. Similarly, the virtual
network operator could decide to implement a new firewall policy in
the virtual network or change the forwarding behavior. The old fabric
would work seamlessly with this new virtual program without
intervention by the physical network operator. In principle, our
compiler could even be used repeatedly to virtualize virtual networks.

\section{Related Work}

A large body of work has explored the design of high-level languages
for \textsc{sdn}
programming~\cite{vmware:nvp,monsanto:pyretic,monsanto:netcore,ferguson:hft,odl:group,onos:intent,voellmy:maple}. Our
work is unique in its focus on the task of engineering efficient
compilers that scale up to large topologies as well as expressive global and
virtual programs.

An early paper by Monsanto et al. proposed the NetCore language and
presented an algorithm for compiling programs based on forwarding
tables~\cite{monsanto:netcore}. Subsequent work by Guha \etal.
developed a verified implementation of NetCore in the Coq proof
assistant~\cite{guha:machine-verified-controllers}. Anderson et
al. developed \netkat as an extension to NetCore and proposed a
compilation algorithm based on manipulating nested conditionals, which
are essentially equivalent to forwarding tables. The correctness of
the algorithm was justified using \netkat's equational axioms, but
didn't handle global programs or Kleene star.  Concurrent
NetCore~\cite{Schlesinger14cnc} grows NetCore with features that
target next-generation \textsc{sdn}-switches.  The original Pyretic
paper implemented an ``reactive microflow interpreter'' and not a
compiler~\cite{monsanto:pyretic}. However later work developed a
compiler in the style of NetCore. \textsc{sdx} uses Pyretic to program
Internet exchange points~\cite{gupta:sdx}. CoVisor develops
incremental algorithms for maintaining forwarding table in the
presence of changes to programs composed using NetCore-like
operators~\cite{covisor}. Recent work by Jose \etal. developed a
compiler based on integer linear programming for next-generation
switches, each with multiple, programmable forwarding
tables~\cite{p4compiler}.

A number of papers in the systems community have proposed mechanisms
for implementing virtual network programs. An early workshop paper by
Casado proposed the idea of network virtualization and sketched an
implementation strategy based on a
hypervisor~\cite{casado:presto}. Our virtual compiler extends this
basic strategy by introducing a generalized notion of a fabric,
developing concrete algorithms for computing and selecting fabrics,
and showing how to compose fabrics with virtual programs in the
context of a high-level language. Subsequent work by Koponen et
al. described VMware's \textsc{nvp} platform, which implements
hypervisor-based virtualization in multi-tenant
datacenters~\cite{vmware:nvp}. Pyretic~\cite{monsanto:pyretic},
CoVisor~\cite{covisor}, and OpenVirteX~\cite{shabibi:ovx} all support
virtualization---the latter at three different levels of abstraction:
topology, address, and control application. However, none of these
papers present a complete description of algorithms for computing the
forwarding state needed to implement virtual networks.

The \fdds used in our local compiler as well as our algorithms for
constructing \netkat automata are inspired by Pous's work on
symbolic \textsc{kat} automata~\cite{pous:symbolickat} and work by
some of the authors on a verification tool
for \netkat~\cite{netkat-automata}. The key differences between this
work and ours is that they focus on verification of programs whereas
we develop compilation algorithms. \bdds have been used for
verification for several
decades~\cite{Akers:1978:BDD:1310167.1310815,Bryant:1986:GAB:6432.6433}. In
the context of networks, \bdds and \bdd-like structures have been used
to optimize access control
policies~\cite{liu:xengine}, \textsc{tcam}s~\cite{tcam-razor}, and to
verify~\cite{khurshid13veriflow} data plane configurations, but our
work is the first to use \bdds to compile network programs.

\section{Conclusion}

This paper describes the first complete compiler for the \netkat
language. It presents a suite of tools that leverage \bdds, graph
algorithms, and symbolic automata to efficiently compile programs in
the \netkat language down to compact forwarding tables
for \textsc{sdn} switches. In the future, we plan to investigate
whether richer constructs such as stateful and probabilistic programs
can be implemented using our techniques, how classic algorithms from
the automata theory literature can be adapted to optimize global
programs, how incremental algorithms can be incorporated into our
compiler, and how the compiler can assist in performing graceful
dynamic updates to network state.

\paragraph*{Acknowledgments.}
The authors wish to thank the anonymous ICFP '15 reviewers, Dexter
Kozen, Shriram Krishnamurthi, Konstantinos Mamouras, Mark Reitblatt,
Alexandra Silva, and members of the Cornell PLDG and DIKU COPLAS
seminars for insightful comments and helpful suggestions. We also wish
to thank the developers of \textsc{gnu} Parallel~\cite{Tange2011a} for
developing tools used in our experiments. Our work is supported by the
National Science Foundation under grants CNS-1111698, CNS-1413972,
CNS-1413985, CCF-1408745, CCF-1422046, and CCF-1253165; the Office of
Naval Research under grants N00014-12-1-0757 and N00014-15-1-2177; and
a gift from Fujitsu Labs.

{
\bibliographystyle{plain}
\balance
\bibliography{main}
}

\end{document}